\documentclass[12pt,a4paper]{iopart}

\newcommand{\eq}[1]{\begin{equation}#1\end{equation}}
\newcommand{\dd}{\mathrm{d}}
\newcommand{\ee}{\mathrm{e}}

\expandafter\let\csname equation*\endcsname\relax
\expandafter\let\csname endequation*\endcsname\relax 

\usepackage{amsmath}
\usepackage{iopams}
\usepackage{amsfonts}
\usepackage{braket}
\usepackage{bbold}
\usepackage{graphics}
\usepackage{graphicx}
\usepackage{leftidx}
\usepackage{psfrag}
\usepackage{color}
\usepackage{colordvi}
\usepackage{bbm}
\usepackage{hyperref}
\hypersetup{colorlinks,bookmarksopen,bookmarksnumbered,
 citecolor=red,
 linkcolor=blue,
 pdfstartview=green,
 urlcolor=purple}

\begin{document}

\title{Entanglement spreading after local and extended excitations in a free-fermion chain}
\author{Viktor Eisler}
\address{
Institute of Theoretical and Computational Physics,
Graz University of Technology, Petersgasse 16, 8010 Graz, Austria}

\begin{abstract}
We study the time evolution of entanglement created by local or extended excitations
upon the ground state of a free-fermion chain. A single particle or hole excitation produces
a single bit of excess entropy for large times and subsystem lengths. In case of a
double hole, some of the coherence between the excitations is preserved and the excess
entropy becomes additive only for large hole separations. In contrast, the coherence is always
lost for particle-hole excitations. Multiple hole excitations on a completely filled
chain are also investigated. We find that for an extended contiguous hole the excess entropy
scales logarithmically with the size, whereas the increase is linear for finite separations
between the holes.

\end{abstract}

%\maketitle

\section{Introduction\label{sec:intro}}

Entanglement properties in one-dimensional quantum systems have been
the topic of intensive research in the last decades \cite{Amico08,CCD09,Laflo16}.
The most emblematic feature uncovered is the existence of an area law,
which bounds the scaling of entanglement entropy in ground states of
local Hamiltonians \cite{ECP10}. Notable exceptions include quantum chains
at criticality \cite{Vidal03}, where the logarithmic violation of the area law shows a
universal character that can be understood via the underlying conformal field
theory (CFT) \cite{CC09}. Such a moderate growth of ground-state entanglement
is the key ingredient behind the classical simulability of quantum chains via
matrix product state methods \cite{Schollwoeck11}.

The area law with eventual logarithmic corrections is, however, not restricted
to the ground-state scenario. Indeed, under some reasonable assumptions, it has been shown
to extend to low-lying excited states of generic local Hamiltonians \cite{Masanes09}.
The simplest case is the one of translational invariant systems possessing well defined
quasiparticle excitations. The entanglement of such few-particle excitations have been
studied in various integrable chains \cite{Pizorn12,Berkovits13,MBSA14,ZR20a,ZR20b},
in the context of quantum field theory \cite{CADFDS18a,CADFDS18b,CADFDS19a,CADFDS19b}, and recently
also in some nonintegrable models \cite{YWPS20,WPSY21}. In the context of free field theory,
few-particle excitations carry a finite amount of excess entanglement, given by a simple binomial
expression via the probabilities of finding the quasiparticles in the subsystem \cite{CADFDS18a,CADFDS18b}.
In particular, the quasiparticle contributions are equal and independent of the momenta.
On the lattice, however, coherence effects between quasiparticles are retained, leading
to corrections that are most dominant for nearby momenta \cite{ZR20a,ZR20b}.

Another class of low-lying excitations that was studied intensively is the one
generated by primary operators in CFT \cite{ABS11,BAS12}.
In the simplest case, these correspond to low-lying excited eigenstates
of the underlying critical quantum chain. More generally, one could consider
excitations created by primaries that are local in space, and thus do not yield
an eigenstate. The time evolution of entanglement after such local operator excitations
was first studied in \cite{NNT14,Nozaki14,HNTW14}, finding a light-cone
propagation and a finite excess entanglement given via the quantum dimension
of the primary operator. The CFT approach was also extended to the case
of multiple excitations \cite{Numasawa16,GHL18,KM20}, finding an additive behaviour
of the excess entropy.

The lattice counterpart of local operator excitations has first been studied
for a critical transverse Ising chain \cite{CR17}, confirming the CFT predictions.
Entanglement evolution after local fermionic excitations, creating a domain-wall
initial state in the ordered phase of the Ising and XY chains were studied in 
\cite{ZGEN15,EME16,EM18,EM20}. The resulting entanglement profiles along the
chain could be understood by a simple quasiparticle picture. Namely, the excess entropy
is determined via the density fraction of the quasiparticles that arrive to the entanglement
cut by a binary expression \cite{EM20}, analogously to the case of quasiparticle eigenstates
\cite{CADFDS18a}. The quasiparticle description of the excess entropy has also been
generalized to the interacting XXZ chain \cite{GE21}.

Here we address the question how the excess entropy changes
when considering multiple local excitations in a free-fermion chain. More precisely,
we consider excitations that can be written as a product of local fermion
creation and annihilation operators and study the ensuing entanglement evolution
via Gaussian techniques \cite{PE09}. Starting with the case of a single particle or
hole excitation, we derive an expression that puts the heuristic quasiparticle picture
obtained in \cite{EM20} on a firm ground. The situation is more complicated
for a double hole excitation, as coherence effects enter the picture and the excess
entropy becomes additive only for very large hole separations.
For finite separations we obtain an asymptotic formula for large times.
Interestingly, we find that particle-hole excitations always lose their coherence
and contribute additively for large times. In case of multiple hole excitations on a
completely filled background, we find that the coherence is essentially preserved
only for an extended contiguous hole, which leads to a logarithmic scaling of the
asymptotic excess entropy with the size of the excitation. In contrast, for a finite
separation between the holes one obtains a linear scaling, as the individual hole
excitations become effectively distinguishable.

The manuscript is structured as follows. The model and the general setting
is introduced in Sec. \ref{sec:model}, along with the main tools required
for subsequent calculations.
The following sections are devoted to the study of entanglement after
single (Sec.~\ref{sec:single}) and double hole (Sec.~\ref{sec:double}),
as well as particle-hole excitations (Sec.~\ref{sec:ph}). Finally, the case
of multiple hole excitations is discussed in Sec. \ref{sec:multi}.
Our conclusions are presented in Sec. \ref{sec:concl}, followed by three
Appendices on various details of the calculations.

\section{Model and setting\label{sec:model}}

We consider an infinite hopping chain given by the Hamiltonian
\eq{
H = -\frac{1}{2}\sum_{n=-\infty}^{\infty} (c^\dagger_n c_{n+1} + c^\dagger_{n+1} c_n)
- \mu \sum_{n=-\infty}^{\infty} c^\dagger_n c_{n} \, ,
\label{H}
}
with fermionic creation/annihilation operators satisfying canonical anticommutation
relations $\{ c_m^\dag,c_n \}=\delta_{mn}$. The Hamiltonian is diagonalized
by a Fourier transform, leading to the dispersion $\omega(q)=-\cos q -\mu$,
where the chemical potential $\mu$ sets the filling of the chain via the
Fermi wavenumber, satisfying $\omega(\pm q_F)=0$. The ground state $\ket{\psi_0}$
of the chain is characterized by its correlation matrix
\eq{
C_{mn} = \bra{\psi_0} c^\dagger_m c^{\phantom{\dagger}}_{n} \ket{\psi_0} =
\frac{\sin\left[q_F(m-n)\right]}{\pi(m-n)} \, ,
}
where the filling is given by $C_{mm}=q_F/\pi$, and all the higher-order
correlation functions factorize according to Wick's theorem.

In the following sections we will consider excitations created upon the
ground state of the chain by acting with an operator $\mathcal{O}$,
supported on a single or a few lattice sites, and then letting the
system evolve, governed by the Hamiltonian in \eqref{H}. 
The time-evolved state can be written as
\eq{
\ket{\psi(t)} = \mathcal{N}_\mathcal{O}^{-1/2} \ee^{-iHt} \mathcal{O} \ket{\psi_0} , \qquad
\mathcal{N}_\mathcal{O} = \bra{\psi_0} \mathcal{O}^\dag \mathcal{O} \ket{\psi_0} ,
}
where the factor $\mathcal{N}_\mathcal{O}$ ensures normalization.
We will focus on a class of excitations that preserve the Gaussianity of the state,
i.e. Wick's theorem also applies to $\mathcal{O} \ket{\psi_0}$, see \ref{app:a}
for details. Since the time evolution is governed by a free-fermion Hamiltonian,
$\ket{\psi(t)}$ is just another Gaussian state characterized by the time-dependent
correlation matrix
\eq{
C_{mn}(t) = \bra{\psi(t)} c^\dagger_m c^{\phantom{\dagger}}_{n} \ket{\psi(t)} .
\label{Cmnt}}
In order to evaluate these matrix elements, one only needs to determine
the effect of the local excitation. Denoting with $C'$ the correlation matrix
immediately after the excitation, the time evolution is obtained via the unitary
transformation
\eq{
C(t) = U^\dag C' \, U \, ,
\label{Ct}}
where the matrix elements of $U$ are given via Bessel functions as
\eq{
U_{mn} = \int_{-\pi}^{\pi} \frac{\dd q}{2\pi} \ee^{it\cos q} \ee^{i(n-m)q}=
i^{n-m}J_{n-m}(t) \, .
\label{U}}

Our goal is to calculate the entanglement entropy of the segment
$A=\left[1,L\right]$ which, owing to the Gaussianity of the problem,
can be obtained via standard correlation-matrix techniques \cite{PE09}. 
Indeed, the entropy can be calculated as
\eq{
S(t) =
%\sum_{k=1}^{L} \left[-\zeta_k(t) \ln \zeta_k(t)- (1-\zeta_k(t)) \ln (1-\zeta_k(t))\right].
\sum_{k=1}^{L} s(\zeta_k(t)) \, , \qquad
s(x) = -x \ln x - (1-x) \ln (1-x)
\label{St}}
from the eigenvalues $\zeta_k(t)$ of the reduced correlation matrix $C_A(t)$,
with matrix elements \eqref{Cmnt} restricted to the interval $m,n \in A$.
In fact, we are rather interested in the excess entropy
\eq{
\Delta S(t) = S(t) - S_0 \, ,
}
where $S_0$ denotes the ground-state entropy. One should stress that 
we are subtracting the entropy \emph{before} the excitation takes place,
thus $\Delta S(0)\ne 0$ and may become negative.
The quantity $\Delta S(t)$ provides information about the excess
entanglement generated by the dynamics with respect to the ground state,
and is studied for a range of single or few-particle excitations in the following sections.

\section{Single hole or particle excitation\label{sec:single}}

We start with the simplest case of a strictly local excitation, a single hole
created by an annihilation operator inserted at site $\ell$ as
\eq{
\ket{\psi_{1h}} = \mathcal{N}^{-1/2} c_\ell \ket{\psi_0} , \qquad
\mathcal{N} = \bra{\psi_0} c_\ell^\dag c^{\phantom{\dag}}_\ell \ket{\psi_0} ,
}
where the normalization factor is given by the density $\mathcal{N}=q_F/\pi$.
As pointed out in the previous section, the only ingredients we need are the
correlation matrix elements $C'_{mn}=\bra{\psi_{1h}} c^\dagger_m c_{n} \ket{\psi_{1h}}$
after the excitation. Applying Wick's theorem one has
\eq{
C'_{mn} = \mathcal{N}^{-1}
\bra{\psi_0} c^\dagger_{\ell} c^\dagger_m c^{\phantom{\dagger}}_{n}c^{\phantom{\dagger}}_{\ell} \ket{\psi_0} =
C_{mn} -  \mathcal{N}^{-1} C_{m \ell} \, C_{\ell n} \, .
\label{Cp1h}}
The matrix $C'$ has a simple form and is determined via the
ground-state correlations. In particular, one has $C'_{m \ell}=C'_{\ell n}=0$, i.e.
the annihilation operator erases all the correlations along the $\ell$-th row and column
of the matrix. Moreover, the change in the density $C'_{mm}$ is non-local, decaying
as a power-law away from the insertion site $m=\ell$, even though the excitation is created
by a local operator. However, the overall particle number decreases by one, $\Tr(C'-C)=-1$,
as it should for a single hole excitation, which follows from the fact that $C^2=C$ for a pure state.

We shall now apply the time evolution on $C'$ as given by \eqref{Ct}. As the first term
in \eqref{Cp1h} is simply the ground-state correlation matrix, the time evolution affects
only the difference term and one has
\eq{
C(t) = C - \Delta C(t) \, ,% \qquad
%\Delta C_{mn}(t) = \mathcal{N}^{-1} \sum_{kj} U^*_{k,m} C_{k \ell} \, C_{\ell j} U_{j,n} \, .
\label{Ct1h}}
where the elements of the difference matrix can be cast as
\eq{
\Delta C_{mn}(t) = \mathcal{N}^{-1} I^*_{m-\ell}(t)I_{n-\ell}(t) \, , \qquad
I_{n-\ell}(t) = \sum_{j} C_{\ell j} U_{j,n} \, .
\label{DC1h}}
Using the integral representations of the matrix elements of $U$ and $C$,
it is easy to show that the propagator is given by
\eq{
I_{n-\ell}(t) =
%\sum_{j} \int_{-q_F}^{q_F} \frac{\dd q}{2\pi} \ee^{iq(j-\ell)}
%\int_{-\pi}^{\pi} \frac{\dd q'}{2\pi} \ee^{it\cos q'} \ee^{iq'(n-j)}=
\int_{-q_F}^{q_F} \frac{\dd q}{2\pi} \ee^{it\cos q} \ee^{iq(n-\ell)} .
\label{int}}
Note that the propagator depends only on the distance $n-\ell$
measured from the location of the excitation. It differs from the
time-evolution operator in \eqref{U} only in the range of integration,
which is now restricted to the Fermi sea. This has the clear physical
meaning that a hole can only be excited from the initially filled modes.

\begin{figure}[htb]
\center
\includegraphics[width=0.7\textwidth]{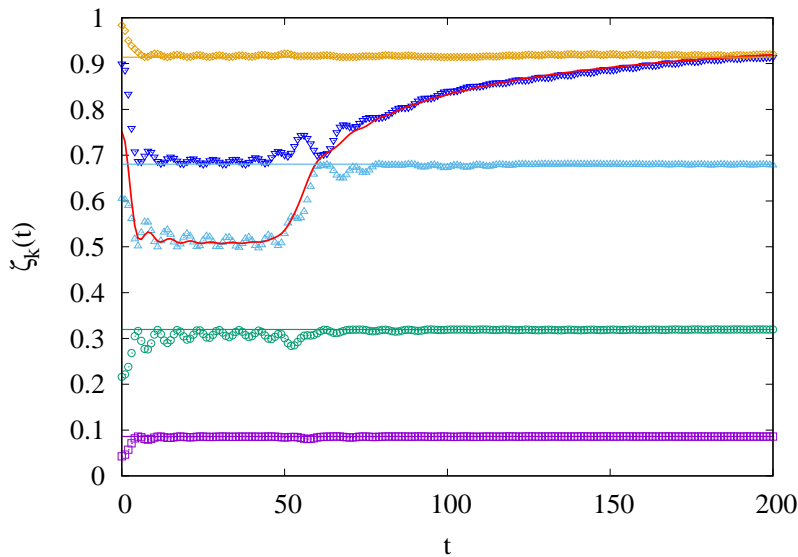}
\caption{Time evolution of the reduced correlation matrix eigenvalues $\zeta_k(t)$
after a single hole excitation with $\ell=0$ at half filling, for an interval of length $L=50$.
The horizontal lines indicate the ground-state eigenvalues. The red solid line shows
$1-\Delta \zeta(t)$, see \eqref{dzetat1h}.}
\label{fig:zetat1h}
\end{figure}

In order to evaluate the entropy, one needs to restrict the correlation
matrix \eqref{Ct1h} to the interval $A$. From the structure of the matrix elements
\eqref{DC1h} one can immediately see, that $\Delta C_A(t)$ is a rank-one
matrix with a single nonzero eigenvalue
\eq{
\Delta \zeta(t) = \mathcal{N}^{-1} \sum_{n=1}^{L} |I_{n-\ell}(t)|^2 .
%= \Tr \Delta C_A (t)
\label{dzetat1h}}
In fact, $\Delta \zeta(t)$ is nothing else but the decrease in the density within the
interval at time $t$ due to the hole excitation.
However, the problem is that the difference matrix does not commute with the
ground-state reduced correlation matrix, $\left[C_A,\Delta C_A(t)\right] \ne 0$,
thus one cannot immediately read off the spectrum of $C_A(t)$.
The eigenvalues $\zeta_k(t)$ in the middle of the spectrum are shown by the
symbols in Fig.~\ref{fig:zetat1h}, for a hole excited at $\ell=0$ next to the left
boundary  of an interval with $L=50$ in a half-filled chain.
One observes that all but one of the eigenvalues move rapidly towards their
ground-state values, indicated by the horizontal lines. The single
anomalous eigenvalue is well captured by $1-\Delta\zeta(t)$, shown by the
red line, up to oscillations which tend to be stronger in the data.
Thus, despite the non-commuting property, the rank-one update $\Delta C_A(t)$
of $C_A$ modifies essentially only one of its
eigenvalues lying exponentially close to one, accounting for the decrease of
density in the subsystem.

%%%%%%%%%%%%%%%%%%%%%%%%%%%%%%%%%%%%%%%%%%%%%%%%%%%%%%%%%%%
%
%
%%%%%%%%%%%%%%%%%%%%%%%%%%%%%%%%%%%%%%%%%%%%%%%%%%%%%%%%%%%

One should remark that the upwards motion of the anomalous eigenvalue is accompanied by avoided
crossings in the spectrum, whenever one of the ground-state eigenvalues is reached.
The first avoided crossing in Fig.~\ref{fig:zetat1h} can be seen at $t \approx 60$ (between light blue
and dark blue symbols) while a second one just starts at $t \approx 200$ (dark blue and yellow),
and the same feature continues towards the upper edge of the spectrum.

\begin{figure}[htb]
\center
\includegraphics[width=0.49\textwidth]{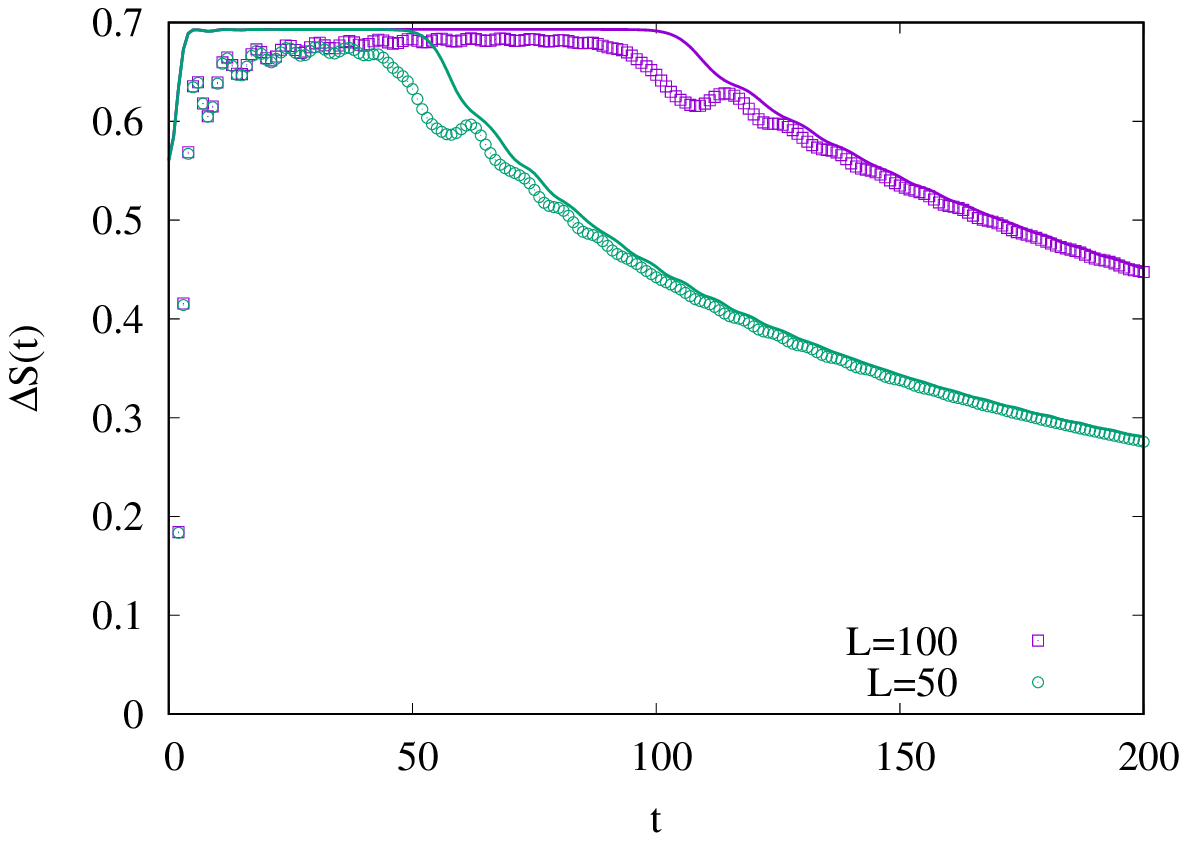}
\includegraphics[width=0.49\textwidth]{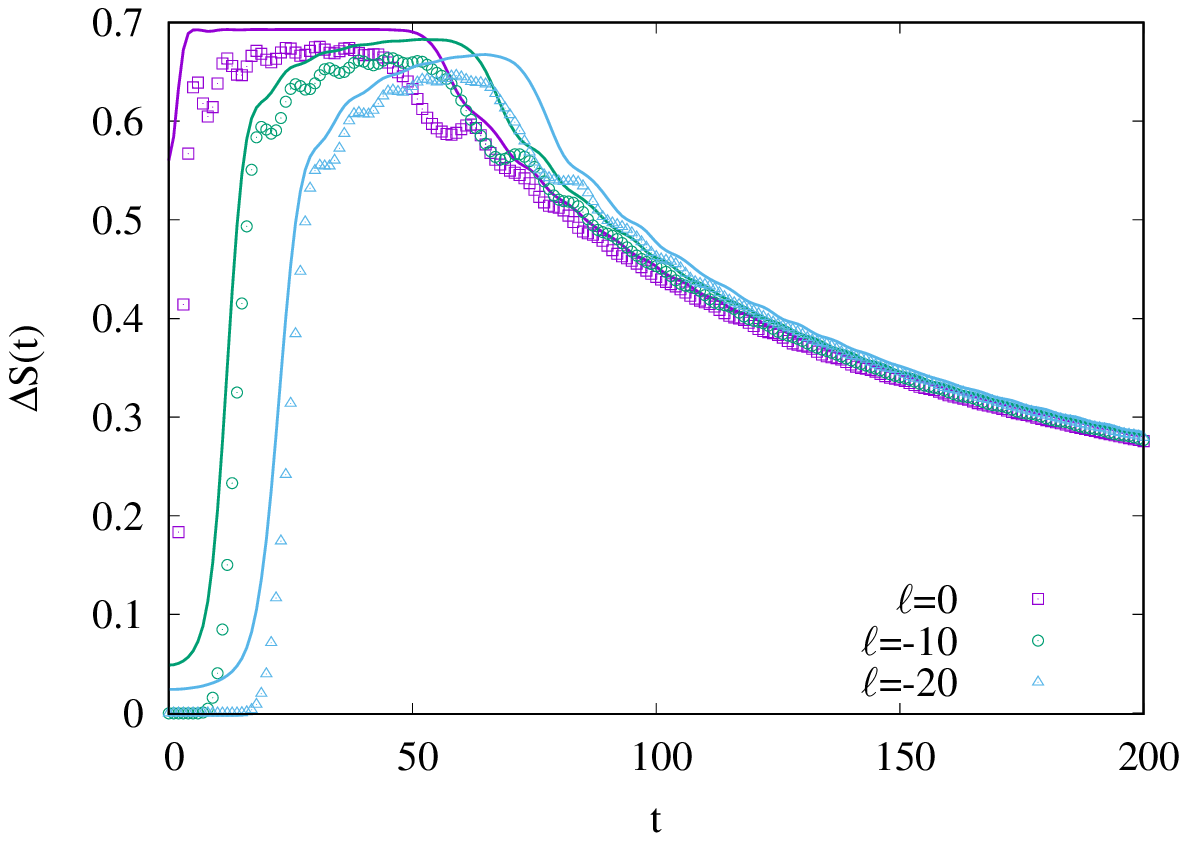}
\caption{Time evolution of the excess entropy $\Delta S(t)$ after a single hole excitation at half filling,
for $\ell=0$ and different interval lengths (left), as well as for various $\ell$ with $L=50$ (right).
The solid lines of matching colors show the approximation $s[\Delta\zeta(t)]$.}
\label{fig:dent1h}
\end{figure}

The above approximate description of the spectra can now be used to understand the
behaviour of the excess entropy, shown on the left of Fig.~\ref{fig:dent1h} for $\ell=0$ and
two different sizes of the interval at half filling. %two interval sizes at half-filling.
Indeed, the change of entropy can mainly be accounted for the
anomalous eigenvalue, and using \eqref{St} one has $\Delta S(t) \approx s[\Delta\zeta(t)]$,
where we used the symmetry $s(x)=s(1-x)$.
This approximation is shown by the lines which feature a plateau region for $0<t<L$,
slightly overestimating the numerical data, followed by a very slow decay for $t>L$
where the approximation becomes increasingly more accurate.
Indeed, this is due to the behaviour of the anomalous eigenvalue in
Fig.~\ref{fig:zetat1h}, which first tends towards the middle of the spectrum,
giving a large contribution to the entropy, while for $t>L$ moves slowly
upwards, yielding the decay of the entropy. Similar features can also be
observed for a hole excitation with $\ell<0$, created at some distance from
the interval, as shown in the right of Fig.~\ref{fig:dent1h}. The main difference
is that the plateau region is shifted towards $|\ell| < t < |\ell| +L$, since the
excitation must first reach the interval, and the plateau is also rounded off.

%%%%%%%%%%%%%%%%%%%%%%%%%%%%%%%%%%%%%%%%%%%%%%%%%%%%%%%%%%%
%
%
%%%%%%%%%%%%%%%%%%%%%%%%%%%%%%%%%%%%%%%%%%%%%%%%%%%%%%%%%%%

For the quantitative understanding of the excess entropy, one has to analyze
the expression in \eqref{dzetat1h}, which is the density fraction of the propagating hole
within the interval. From simple symmetry arguments one finds that
for $|\ell| \ll t \ll |\ell| +L$ one has $\Delta \zeta(t) \to 1/2$, since half of the density
travels to the right from the initial hole location. This immediately yields
that the height of the entropy plateau is given by $\ln 2$. Furthermore, one can
even take into account the dispersion of the fermionic modes in a semi-classical
picture. This is equivalent to calculating the stationary-phase approximation of \eqref{dzetat1h}
using the integral representation of the propagator \eqref{int}. As shown in \ref{app:b},
for $t, L\gg 1$ and $\ell \le 0$, this leads to the following expression
\eq{
\Delta\zeta(t) \simeq \int_{-q_F}^{q_F} \frac{\dd q}{2 \, q_F}
\Theta(v_qt-|\ell|-1/2)\Theta(L+|\ell|+1/2-v_qt) \, ,
\label{dzetatsc}}
where $v_q=\frac{\dd \omega}{\dd q}=\sin q$ are the single-particle velocities
and $\Theta(x)$ is the Heaviside step function.
This has the obvious semi-classical interpretation that the change of density is given by the
fraction of the propagating modes located within the interval. Note that the $1/2$
distance correction in the arguments of the step functions is due to the fact,
that the initial hole density is not sharply localized at site $\ell$. This correction
only becomes important for small $|\ell|$ and at short times.

%%%%%%%%%%%%%%%%%%%%%%%%%%%%%%%%%%%%%%%%%%%%%%%%%%%%%%%%%%%
%
\begin{figure}[htb]
\center
\includegraphics[width=0.7\textwidth]{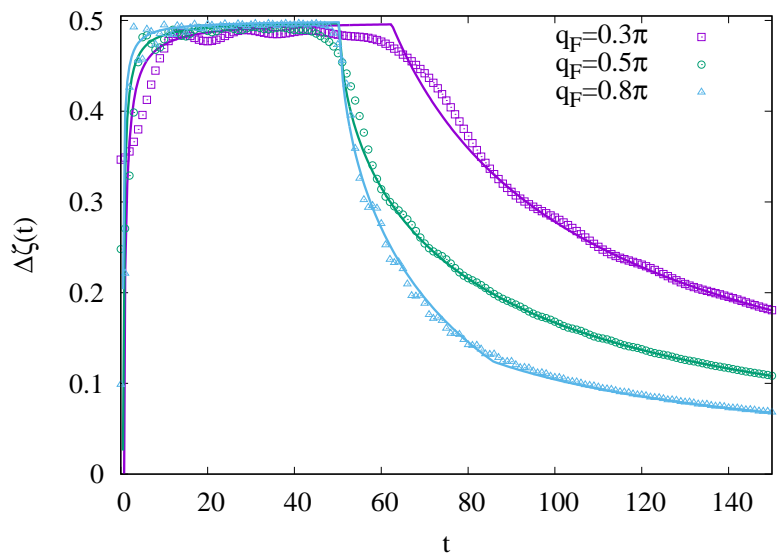}
\caption{Time evolution of $\Delta \zeta(t)$ after a single hole excitation with $\ell=0$ 
at various fillings and $L=50$ (symbols), compared to the semi-classical approximation
in \eqref{dzetatsc} (lines).}
\label{fig:dzetatsc}
\end{figure}
%
%%%%%%%%%%%%%%%%%%%%%%%%%%%%%%%%%%%%%%%%%%%%%%%%%%%%%%%%%%%

To check the validity of \eqref{dzetatsc}, in Fig.~\ref{fig:dzetatsc} we have
calculated $\Delta \zeta(t)$ for $\ell=0$ and various fillings, finding a good agreement
with the semi-classical formula up to oscillations. One also observes a
change of behaviour below and above half filling. Indeed, for $q_F<\pi/2$
one has $v_{max}=v_{q_F}$, and $\Delta \zeta(t)$ starts to decay at $v_{q_F}t \approx L$.
In contrast, for $q_F>\pi/2$ one has $v_{max}=1$, and the decay starts already
at $t \approx L$. However, there is a clear change of behaviour also at $v_{q_F}t \approx L$,
where the modes $\pi/2 < q < q_F$ having a pair $\pi-q$ with the same
velocity, are depleted. This changes the slope of decay and is observed as a
cusp in the semi-classical approximation for $q_F=0.8\pi$ in Fig.~\ref{fig:dzetatsc}.
The numerical data, however, interpolates smoothly between these two regimes.
It is also interesting to have a look at the asymptotic decay. 
For $v_{q_F}t \gg L$ one has
\eq{
\Delta \zeta(t) \approx \frac{1}{2 \, q_F}\arcsin \frac{L}{t} \approx \frac{L}{2\, q_F \, t} \, ,
}
which yields for the excess entropy
\eq{
\Delta S(t) \approx \frac{L}{2\, q_F \, t} \left(\ln \frac{2\, q_F \, t}{L}+1\right) .
%+ \mathcal{O}(t^{-2})
\label{DSas1h}}
Thus $\Delta S(t)$ receives a logarithmic correction and its decay is slower than algebraic.

Finally, it should be stressed that the above described relation between
entropy and excess density improves for increasing filling. Indeed, in
the case of complete filling $q_F = \pi$, the correlation matrix becomes
$C_{ij}=\delta_{ij}$, and thus $\left[C_A,\Delta C_A(t)\right] = 0$.
Hence the formula $\Delta S(t) = s[\Delta\zeta(t)]$ for the excess entropy
becomes exact in this case.

To conclude this section, we also discuss the single particle excitation given by
\eq{
\ket{\psi_{1p}} = \bar{\mathcal{N}}^{-1/2} c^\dag_\ell \ket{\psi_0} , \qquad
\bar{\mathcal{N}} = 1-\mathcal{N} .
}
The correlation matrix immediately after the excitation reads
\eq{
C'_{mn} = \bar{\mathcal{N}}^{-1}
\bra{\psi_0} c^{\phantom{\dagger}}_{\ell} c^\dagger_m c^{\phantom{\dagger}}_{n}c^\dagger_{\ell} \ket{\psi_0} =
C_{mn} +  \bar{\mathcal{N}}^{-1} \bar C_{m \ell} \bar C_{\ell n} \, ,
\label{Cp1p}}
where $\bar C_{mn}=\delta_{mn}-C_{mn}$ is the ground-state
correlation matrix after a particle-hole transformation.
After time evolution one then has
\eq{
C(t) = C + \Delta C(t),
}
where
\eq{
\Delta C_{mn}(t) = \bar{\mathcal{N}}^{-1} \bar I^*_{m-\ell}(t) \bar I_{n-\ell}(t) \, , \qquad
\bar I_{n-\ell}(t) = \sum_{j} \bar C_{\ell j} U_{j,n} \, .
}
Here the propagator is given by the integral
\eq{
\bar I_{n-\ell}(t) =
\int\displaylimits_{q\in \bar F} \frac{\dd q}{2\pi} \ee^{it\cos q} \ee^{iq(n-\ell)} \, ,
\label{barint}}
where $\bar F = \left[q_F,\pi \right] \cup \left[-\pi, -q_F \right]$ is the complement
of the Fermi sea $F$, reflecting the fact that a single particle excitation can only create modes
that were initially empty. It is easy to see, that the particle excitation with filling $q_F/\pi$ is trivially
related to the hole excitation with filling $1-q_F/\pi$ via a particle-hole transformation.

\section{Double hole excitation\label{sec:double}}

Next we consider a double hole excitation given by
\eq{
\ket{\psi_{2h}} = \mathcal{N}_2^{-1/2} c_{\ell_2}c_{\ell_1} \ket{\psi_0} , \qquad
\mathcal{N}_2 = \bra{\psi_0} c_{\ell_1}^\dag c_{\ell_2}^\dag c_{\ell_2}^{\phantom{\dag}} c_{\ell_1}^{\phantom{\dag}} \ket{\psi_0}=
\mathcal{N}^2 - C^2_{\ell_1 \ell_2} \, ,
}
where $\ell_1 \ne \ell_2$. The correlation matrix immediately after the excitation
can again be evaluated by Wick's theorem and reads
\eq{
C'_{mn}=
C_{mn} -  \mathcal{N}_2^{-1} \left[
\mathcal{N}(
C_{m \ell_1} C_{\ell_1 n}+
C_{m \ell_2} C_{\ell_2 n}) -
C_{\ell_1 \ell_2}(
C_{m \ell_1}  C_{\ell_2 n} +
C_{m \ell_2} C_{\ell_1 n})
\right] .
}
%
%It is easy to check that $C'_{\ell_1 n} = C'_{\ell_2 n} = C'_{m \ell_1} = C'_{m \ell_2} =0$
%for arbitrary $m$ and $n$, as it should.
After time evolution one has again the form \eqref{Ct1h} with the difference matrix given by
\eq{
\begin{split}
\Delta C_{mn}(t) = %a%\mathcal{N}/\mathcal{N}_2
%&\left[I_{m-\ell_1}^*(t)I_{n-\ell_1}(t) + I_{m-\ell_2}^*(t)I_{n-\ell_2}(t)\right] \\
%+ \,\, b%- C_{\ell_1 \ell_2}/\mathcal{N}_2 
%&\left[ I_{m-\ell_1}^*(t) I_{n-\ell_2}(t) + I_{m-\ell_2}^*(t)I_{n-\ell_1}(t) \right] .
\mathcal{N}^{-1}\sum_{\alpha,\beta=1}^{2} A_{\alpha\beta} I^*_{m-\ell_\alpha}(t)I_{n-\ell_\beta}(t) \, , \qquad
A = \left(
\begin{array}{cc}
a & b \\
b & a
\end{array}
\right),
\label{DC2h}
\end{split}}
where we have defined $a = \mathcal{N}^2/\mathcal{N}_2$ and
$b = - \mathcal{N}C_{\ell_1 \ell_2}/\mathcal{N}_2$. The structure of \eqref{DC2h}
suggests that we have now to deal with a rank-two update
of the correlation matrix. However, one cannot immediately
write down the eigenvalues of $\Delta C(t)$ since the basis states
$I_{n-\ell_\alpha}(t)$ with $\alpha=1,2$ are not orthonormal.
Nevertheless, introducing the notation
\eq{
\lambda_\alpha = \mathcal{N}^{-1}\sum_{n=1}^{L}|I_{n-\ell_\alpha}(t)|^2, \qquad
\Delta =  \mathcal{N}^{-1} \sum_{n=1}^{L}I^*_{n-\ell_2}(t) I^{\phantom{*}}_{n-\ell_1}(t) \, ,
\label{lamdel}}
as well as $\lambda_{\pm} = (\lambda_1 \pm \lambda_2)/2$,
it is shown in \ref{app:c} that the two nonzero eigenvalues of $\Delta C(t)$
can be obtained as
\eq{
\Delta\zeta_{1,2}  = a \lambda_+ + b \, \mathrm{Re} \, \Delta \pm
\sqrt{(a \lambda_+ + b \, \mathrm{Re} \, \Delta)^2 - (a^2-b^2)(\lambda_1\lambda_2-|\Delta|^2)}  \, .
\label{dzetat2h}}
Note that we have suppressed the argument $t$ for notational simplicity.
It should be stressed that the eigenvalues now depend, apart from the densities
$\lambda_{1,2}$, also on the complex overlap $\Delta$ in \eqref{lamdel}. In the special
case $C_{\ell_1\ell_2}=0$ one has $a=1$ and $b=0$ and thus \eqref{dzetat2h}
simplifies to
\eq{
\Delta\zeta_{1,2}  = \lambda_+ \pm \sqrt{\lambda^2_- + |\Delta|^2}\, .
\label{dzetat2hb}}

\begin{figure}[htb]
\center
\includegraphics[width=0.49\textwidth]{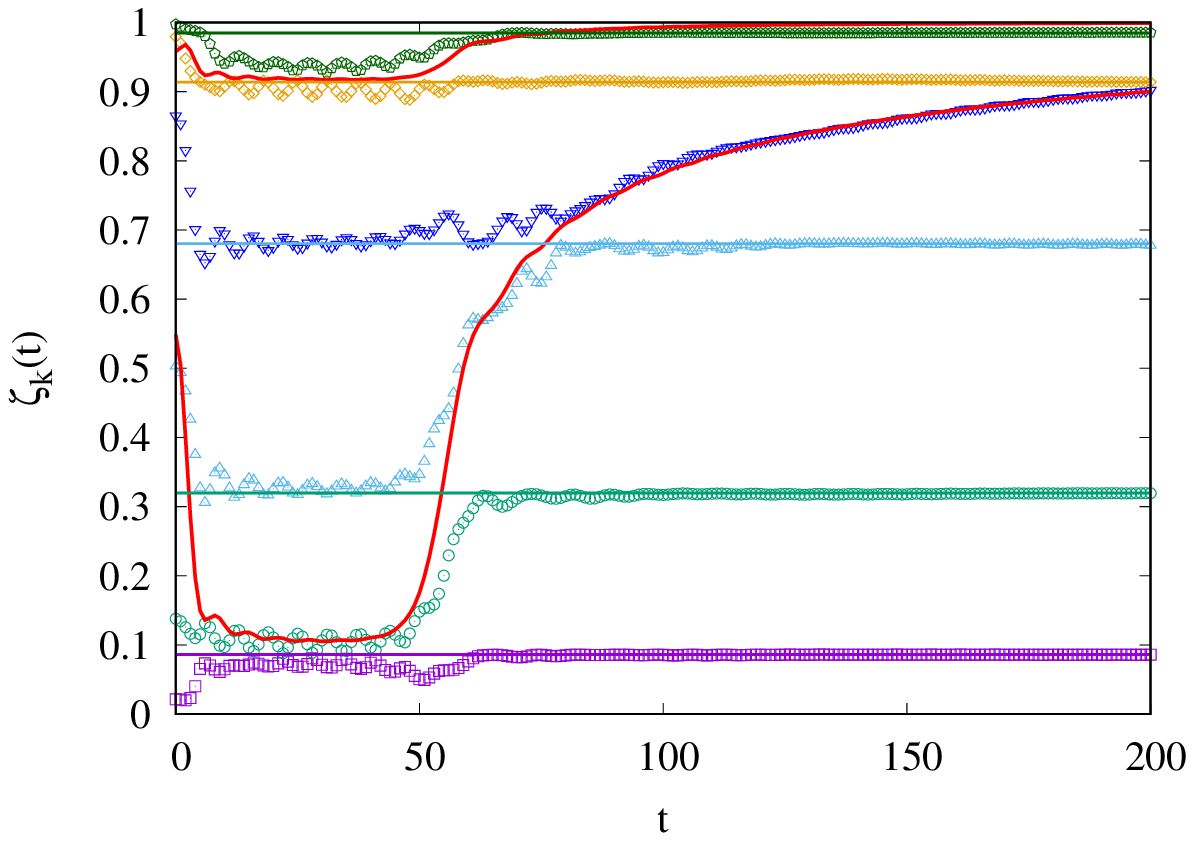}
\includegraphics[width=0.49\textwidth]{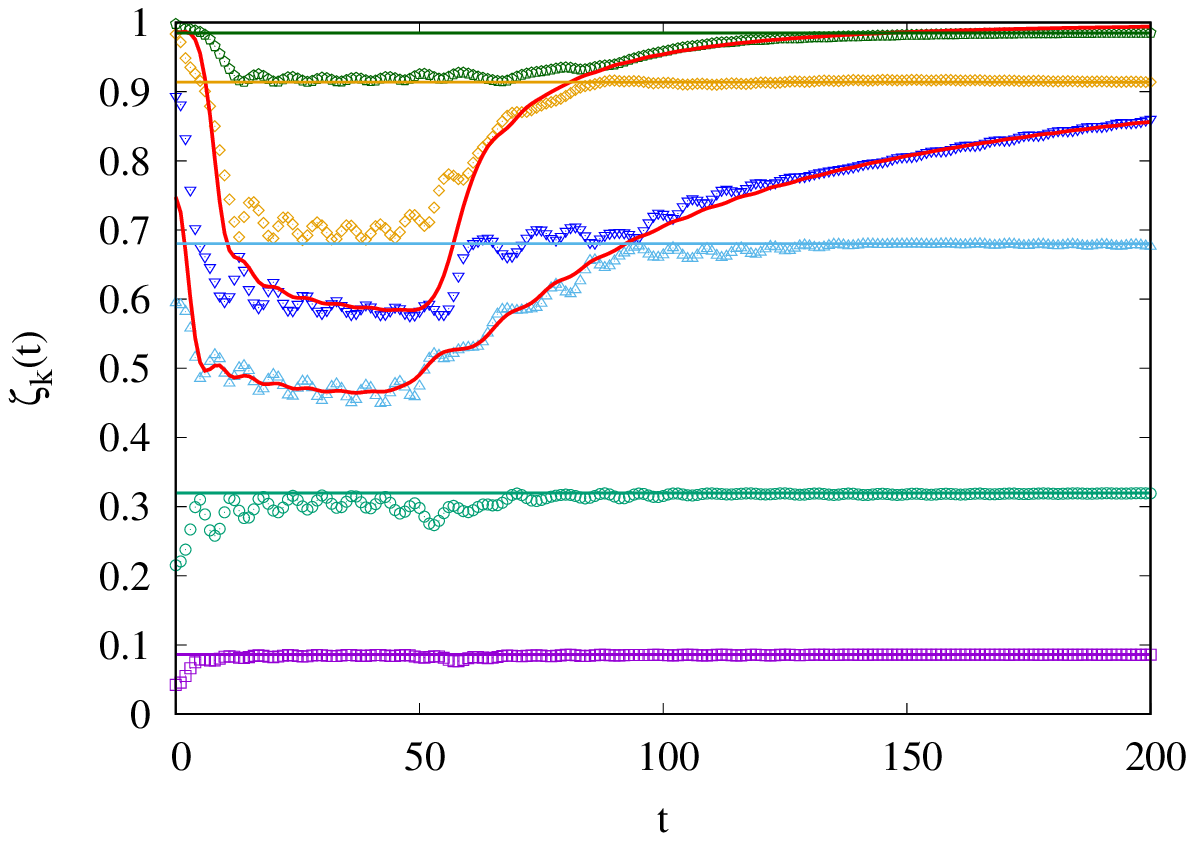}
\caption{Time evolution of the $\zeta_k(t)$ after a double hole excitation
with $\ell_1=0$ and $\ell_2=-1$ (left) as well as $\ell_2=-5$ (right), for an interval of length $L=50$
at half filling. The horizontal lines indicate the ground-state eigenvalues.
The red solid lines show $1-\Delta \zeta_{1,2}$, see \eqref{dzetat2h}.}
\label{fig:zeta2h}
\end{figure}

Analogously to the case of a single hole excitation, we now expect that
two anomalous eigenvalues should appear around $1-\Delta \zeta_{1,2}$
in the spectrum $\zeta_k(t)$. This is illustrated in Fig.~\ref{fig:zeta2h}
for two different separations between the holes. On the left we have
$\ell_1=0$ and $\ell_2=-1$, such that the holes are located at neighbouring
sites next to the boundary of the interval. For $t<L$ one observes that
there is indeed an eigenvalue well described, up to oscillations, by the smaller
of the two $1-\Delta \zeta_{1,2}$, whereas the larger one lies between two
$\zeta_k(t)$ that tend to converge toward it from both sides. The decay regime $t>L$
is dominated by the smaller anomalous eigenvalue, which moves slowly
upwards through the spectrum. On the right of Fig.~\ref{fig:zeta2h} we
also show the case $\ell_1=0$ and $\ell_2=-5$, where both $1-\Delta \zeta_{1,2}$
seem to give a better description of the anomalous eigenvalues,
featuring similar avoided crossings as in the single-hole case in Fig.~\ref{fig:zetat1h}.

%%%%%%%%%%%%%%%%%%%%%%%%%%%%%%%%%%%%%%%%%%%%%%%%%%%%%%%%%%%
%
%
%%%%%%%%%%%%%%%%%%%%%%%%%%%%%%%%%%%%%%%%%%%%%%%%%%%%%%%%%%%

%%%%%%%%%%%%%%%%%%%%%%%%%%%%%%%%%%%%%%%%%%%%%%%%%%%%%%%%%%%
%
\begin{figure}[htb]
\center
\includegraphics[width=0.49\textwidth]{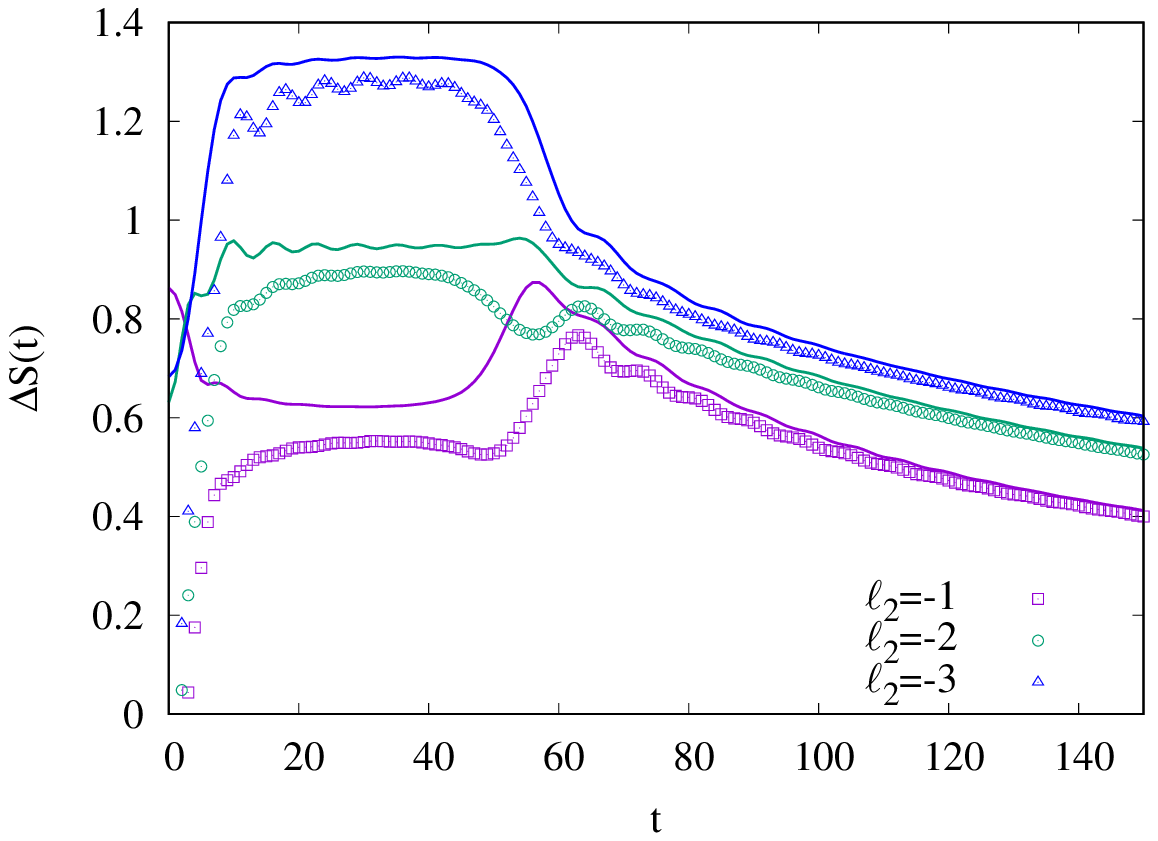}
\includegraphics[width=0.49\textwidth]{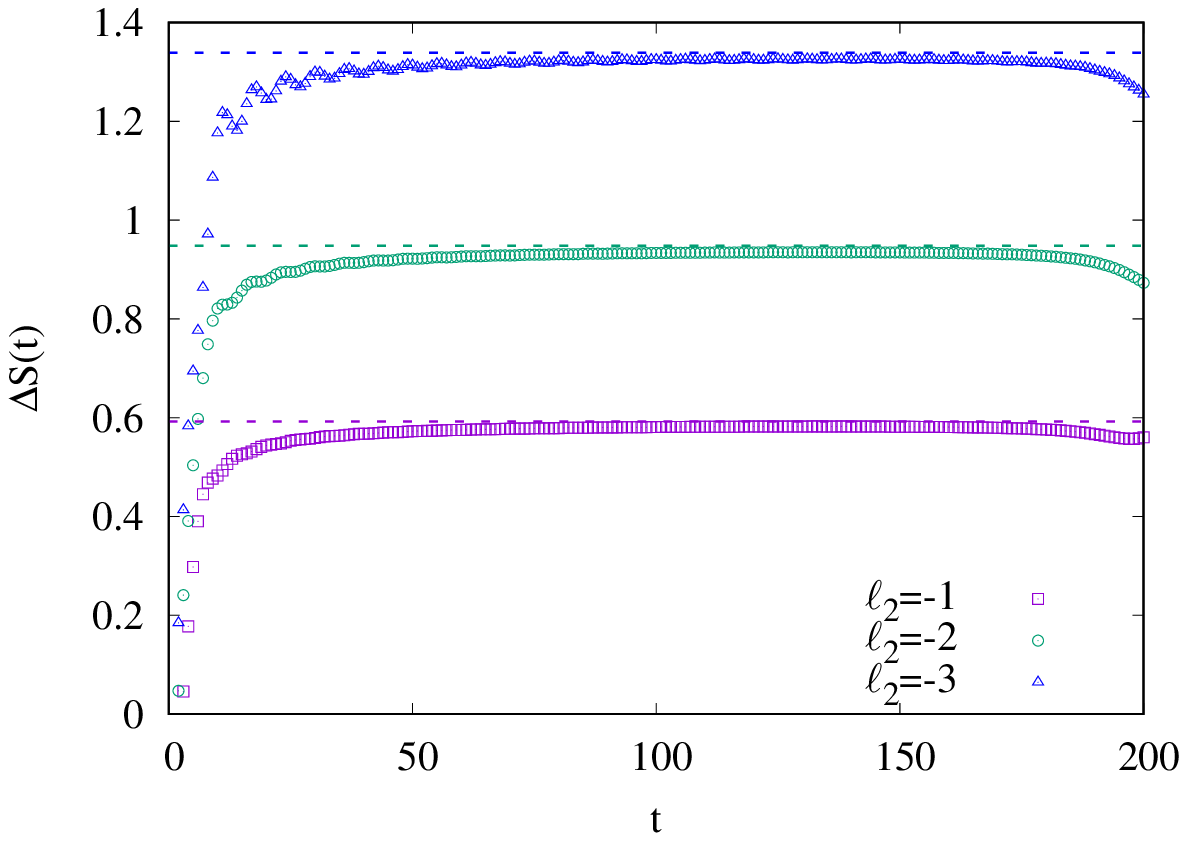}
\caption{Excess entropy $\Delta S(t)$ for double hole excitations
with $\ell_1=0$ and various $\ell_2$, for an interval of size $L=50$ (left)
as well as $L=200$ (right) at half filling.
The solid lines indicate the approximation \eqref{DS2h}.
The horizontal dashed lines indicate the asymptotic result using the
eigenvalues \eqref{dzetasp}.}
\label{fig:dent2h}
\end{figure}
%
%%%%%%%%%%%%%%%%%%%%%%%%%%%%%%%%%%%%%%%%%%%%%%%%%%%%%%%%%%%

The behaviour of the excess entropy is shown in Fig.~\ref{fig:dent2h} for $\ell_1=0$
and various $\ell_2$. One expects that the entropy is now approximated by
\eq{
\Delta S(t) \approx s(\Delta\zeta_1) + s(\Delta\zeta_2) \, ,
\label{DS2h}}
which is shown by the solid lines for $L=50$ on the left of Fig.~\ref{fig:dent2h}.
While the approximation seems to improve for increasing separations between
the holes, \eqref{DS2h} always overestimates the excess entropy and fails especially 
for short times $t \ll L$. This is most prominent in case of $\ell_2=-1$, which could already
be anticipated from the behaviour of $\Delta\zeta_{1,2}$ in Fig.~\ref{fig:dent2h}.
Nevertheless, one also observes that the ansatz \eqref{DS2h} improves for larger times,
and it correctly gives the height of the plateau in the asymptotic regime $|\ell_1|,|\ell_2| \ll t \ll L$.
This is illustrated on the right of Fig.~\ref{fig:dent2h}, where the dashed lines indicate
the ansatz \eqref{DS2h} with the $\Delta\zeta_{1,2}$ evaluated via stationary-phase analysis.
Indeed, for the densities one obtains $\lambda_{1,2} \simeq 1/2$ as for the single hole,
whereas the overlap in \eqref{lamdel} is evaluated in \ref{app:b} and yields
\eq{
%\lambda_1=\lambda_2 = \frac{\mathcal{N}}{2} \, , \qquad
\Delta \simeq \int_{0}^{q_F} \frac{\dd q}{2 q_F} \ee^{iq(\ell_2-\ell_1)}=
%\frac{C_{\ell_1\ell_2}}{2\mathcal{N}} + i \frac{\cos q_F(\ell_1-\ell_2)-1}{2q_F(\ell_1-\ell_2)} \, .
\frac{\sin \left[q_F(\ell_1-\ell_2)\right]}{2q_F(\ell_1-\ell_2)} +
i \frac{\cos \left[q_F(\ell_1-\ell_2)\right]-1}{2q_F(\ell_1-\ell_2)} \, .
\label{delsp}}
%
%Using $a \lambda_+ + b \, \mathrm{Re}(\Delta) = 1/2$,
Substituting into \eqref{dzetat2h}, one obtains for the asymptotic eigenvalues
\eq{
%\Delta S =  s(\Delta \zeta_1) + s(\Delta \zeta_2) \, ,
%\qquad
%\Delta \zeta_{1,2} = \frac{1}{2} \pm \frac{|C'_{\ell_1\ell_2}|}{\sqrt{\mathcal{N}_2}} \, .
\Delta \zeta_{1,2} = \frac{1}{2} \pm \frac{\mathcal{N}|\mathrm{Im} (\Delta) |}{\sqrt{\mathcal{N}_2}} \, .
\label{dzetasp}}
Interestingly, the plateau height is not monotonously increasing with
the distance $|\ell_1-\ell_2|$. In particular, at half filling one has $\mathrm{Im} (\Delta)=0$
for $\mod(|\ell_1-\ell_2|,4)=0$ and thus one obtains the maximal excess entropy $\Delta S =2\ln 2$.
In general, $\Delta S$ also converges towards this maximum at large separations
$|\ell_1-\ell_2|\gg 1$, since from \eqref{delsp} one has $\Delta \to 0$
as the coherence between the two holes is lost and they become effectively independent.
%In other words, the entanglement contributions from the two excitations
%become additive.

\section{Particle-hole excitations\label{sec:ph}}

Our next example is the particle-hole excitation
\eq{
\ket{\psi_{ph}} = \mathcal{N}_{ph}^{-1/2}
(c^\dag_{\ell_2}c^{\phantom{\dag}}_{\ell_1}-C_{\ell_2\ell_1}) \ket{\psi_0} , \qquad
\mathcal{N}_{ph} = \mathcal{N}\bar{\mathcal{N}} \, ,
\label{phi0ph}}
where, in order to ensure orthogonality $\braket{\psi_0|\psi_{ph}}=0$,
we subtracted the ground-state expectation value from the operator.
The particle-hole excitation defined this way still leads to a Gaussian state,
as verified in \ref{app:a}. The correlation matrix after the excitation then reads
\eq{
C'_{mn}=
C_{mn} - 
\mathcal{N}^{-1} C_{m \ell_1} C_{\ell_1 n} +
\bar{\mathcal{N}}^{-1} \bar{C}_{m \ell_2} \bar{C}_{\ell_2 n} \, .
\label{Cpph}}
Thus the contributions from a single hole \eqref{Cp1h} as well as from
a single particle \eqref{Cp1p} become additive, entering the expression with
different signs. After time evolution this yields
\eq{
C_{mn}(t) = C_{mn} - \mathcal{N}^{-1} I_{m-\ell_1}^*(t)I_{n-\ell_1}(t)
+\bar{\mathcal{N}}^{-1} \bar I_{m-\ell_2}^*(t) \bar I_{n-\ell_2}(t) \, .
\label{Ctph}}
Following the arguments applied for a single excitation,
one expects two anomalous eigenvalues to appear around
$1-\lambda_1$ and $\bar \lambda_2$, respectively, where
\eq{
\lambda_1 = \mathcal{N}^{-1}\sum_{n=1}^{L}|I_{n-\ell_1}(t)|^2 \, , \qquad
\bar \lambda_2 = \mathcal{\bar N}^{-1}\sum_{n=1}^{L}|\bar I_{n-\ell_2}(t)|^2 \, .
}
Consequently, the excess entropy should also be approximately additive
\eq{
\Delta S \approx s(\lambda_1) + s(\bar \lambda_2) \, .
\label{DSph}}

\begin{figure}[htb]
\center
\includegraphics[width=0.49\textwidth]{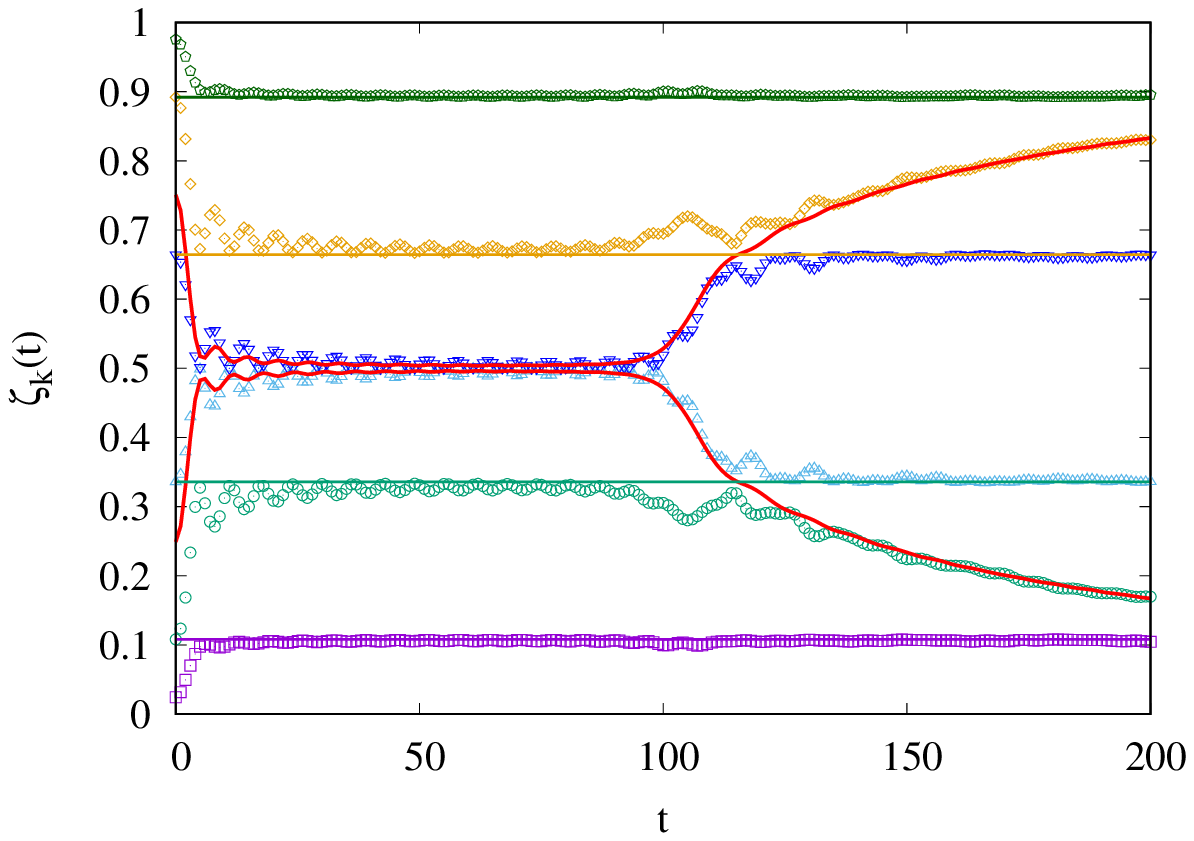}
\includegraphics[width=0.49\textwidth]{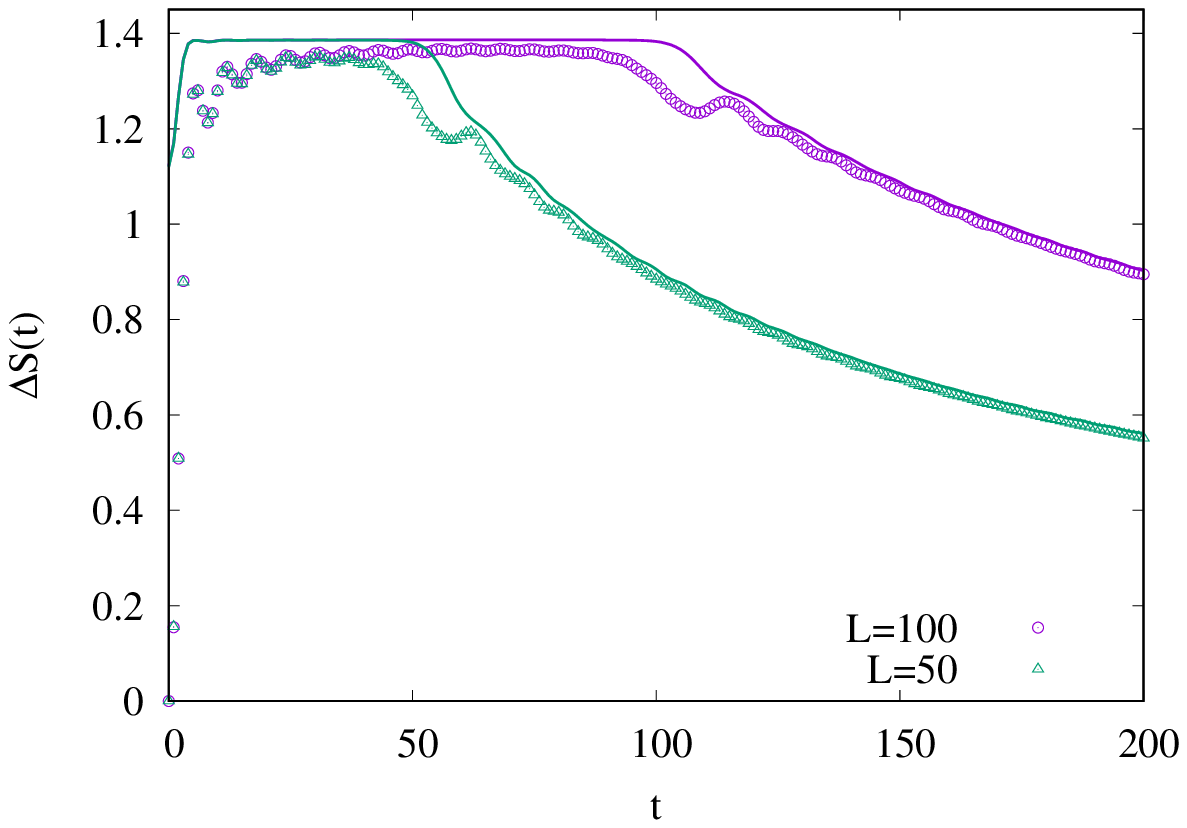}
\caption{
Left: Time evolution of the spectrum $\zeta_k(t)$ after a particle-hole excitation
with $\ell_1=\ell_2=0$ and $L=100$ at half filling. The horizontal lines indicate the ground-state
eigenvalues. The red solid lines show $1-\lambda_1$ and $\bar \lambda_2$, respectively. 
Right: time evolution of the excess entropy $\Delta S(t)$ for $L=50$ and $L=100$, with
the ansatz \eqref{DSph} shown by the solid lines.}
\label{fig:zetaentph}
\end{figure}

The above ansatz is checked against numerical results in Fig.~\ref{fig:zetaentph}
for $\ell_1=\ell_2=0$ at half filling. On the left one can clearly see that the
anomalous eigenvalues are well captured in the $\zeta_k(t)$  spectrum.
The excess entropies are shown on the right of Fig.~\ref{fig:zetaentph} for two
interval sizes, showing convergence
towards the predicted plateau with $\Delta S = 2\ln 2$. The results are similar
for $\ell_2 < \ell_1 < 0$, with the emergence of an intermediate plateau around
$\Delta S = \ln 2$ in the regime $|\ell_1| \ll t  \ll |\ell_2|$, in case of larger separations.
However, in contrast to the double hole excitation, the asymptotic excess entropy
for $|\ell_2| \ll t \ll L$ is always given by $\Delta S = 2\ln 2$.

%%%%%%%%%%%%%%%%%%%%%%%%%%%%%%%%%%%%%%%%%%%%%%%%%%%%%%%%%%%
%
%
%%%%%%%%%%%%%%%%%%%%%%%%%%%%%%%%%%%%%%%%%%%%%%%%%%%%%%%%%%%

In order to understand this, let us first remark that the two difference terms in
\eqref{Ctph} actually do not commute exactly. In fact, one could proceed similarly to
the double hole case, treating the rank-2 update matrix as in \eqref{DC2h}.
The eigenvalue problem could be dealt with in a very similar fashion, by introducing the
overlap between the particle and hole propagator 
\eq{
\Delta' =  (\mathcal{N \bar N})^{-1/2}
\sum_{n=1}^{L}\bar I^*_{n-\ell_2}(t) I^{\phantom{*}}_{n-\ell_1}(t) \, .
}
However, this overlap decays very quickly towards zero, which can be understood
from a stationary-phase argument. Indeed, using the definitions \eqref{int} and \eqref{barint},
the phase of the resulting double integral could be made stationary by choosing equal
momenta, see \ref{app:b}. However, since the domains of the two integrals are now complementary
in momentum space, the stationary phase condition cannot be satisfied.
This explains why the coherence is not preserved for large times in case of a
particle-hole excitation, which thus effectively behaves as two independent local excitations.

Finally, one should remark that the plateau value of the excess entropy for $\ell_1=\ell_2=\ell$
could also be obtained via the CFT approach to local operator excitations \cite{NNT14,Nozaki14,HNTW14}.
Indeed, using the standard bosonization procedure \cite{Giamarchi}, the operator applied to the
ground state in \eqref{phi0ph} can be rewritten, up to an overall normalization factor,
in terms of the chiral boson fields $\phi$ and $\bar \phi$ as
\eq{
c^\dag_{\ell}c^{\phantom{\dag}}_{\ell}-\mathcal{N} \,\, \rightarrow \,\,
i \partial \phi - i  \bar \partial \bar \phi +
\ee^{2iq_F\ell} \ee^{-i\phi+i\bar\phi} + \ee^{-2iq_F\ell} \ee^{i\phi-i\bar\phi} \, .
\label{boson}}
Here the first two terms are the chiral (right- and left-moving) currents,
corresponding to primary fields with conformal dimensions $(h,\bar h)=(1,0)$
and $(0,1)$, respectively. The last two terms correspond to Umklapp processes,
removing a fermion at the right edge of the Fermi sea and placing it back at the
left edge or vice-versa, and they both have conformal dimensions $(1/2,1/2)$.
Thus the operator in \eqref{boson} is a sum of chiral and non-chiral primaries
with the same scaling dimension $h+\bar h =1$. This is exactly the situation
considered in \cite{ZC20}, finding the result that the excess entropy is simply given
by the Shannon entropy calculated from the weights of the various operators.
However, since \eqref{boson} is an equally weighted sum, one recovers immediately
the result $\Delta S =2 \ln 2$.

%Choosing the definition as in \eqref{Ct1h}, one has for the
%time-evolved difference matrix
%
%\eq{
%\Delta C_{mn}(t) = 
%\mathcal{N}^{-1} I_{\ell_1}^*(m)I_{\ell_1}(n)
%-\bar{\mathcal{N}}^{-1} \bar I_{\ell_2}^*(m) \bar I_{\ell_2}(n) \, .
%}
%
%Note that this is again a rank-2 update and has a similar structure as \eqref{DC2h}
%for the double hole, however with unequal diagonal and vanishing offdiagonal elements.
%Introducing
%
%\eq{
%\bar \lambda_2 = \mathcal{\bar N}^{-1}\sum_{n=1}^{L}|\bar I_{n-\ell_2}(t)|^2, \qquad
%
%\label{lamdelph}}
%
%and following a similar procedure as in \ref{app:c}, one can easily show
%that the nontrivial eigenvalues of the difference matrix read
%
%\eq{
%\Delta\zeta_{1,2}  = \frac{\lambda_1-\bar\lambda_2}{2} \pm
%\sqrt{\left(\frac{\lambda_1+\bar\lambda_2}{2}\right)^2-|\Delta' |^2}  \, .
%\label{dz12ph}}
%
%Before comparing to the spectra $\zeta_k(t)$, one should remark that the
%two eigenvalues in \eqref{dz12ph} have different signs, with $\Delta\zeta_1>0$
%and $\Delta\zeta_2<0$ corresponding to the hole and particle contributions, respectively.
%Following the logic 

\section{Multiple hole excitations\label{sec:multi}}

In the last part we shall consider the case of multiple hole excitations.
For the sake of simplicity, we restrict our attention to the case of a completely
filled chain, $q_F=\pi$, and the excited state reads
\eq{
\ket{\psi_{X}} = \prod_{\ell \in X} c_{\ell} \ket{\psi_0} ,}
where $X$ is an index set labeling the empty sites.
Indeed, for a such a simple initial state the difference matrix can be found explicitly as
\eq{
\Delta C_{mn}(t) = \sum_{\ell \in X} I_{m-\ell}^*(t)I_{n-\ell}(t) \, ,
\label{DCK}}
where the propagator $I_{n-\ell}(t)=i^{n-\ell}J_{n-\ell}(t)$ is given via the Bessel function.
Furthermore, due to $C_{mn}=\delta_{mn}$, the spectrum of the reduced correlation
matrix is given exactly by $\zeta_k(t)=1-\Delta\zeta_k(t)$, where $\Delta\zeta_k(t)$
are the eigenvalues of \eqref{DCK} reduced to the subsystem, $m,n \in A$.
Note also that $S_0=0$ and thus $\Delta S(t)=S(t)$.

In order to understand the entanglement evolution from such a multiple hole excitation,
we shall apply a duality argument. First of all note, that thanks to the property
\eq{
%\sum_{\ell=-\infty}^{\infty} I_{\ell}^*(m)I_{\ell}(n)=
\sum_{\ell=-\infty}^{\infty}J_{m-\ell}(t) J_{n-\ell}(t)=
\delta_{mn} \, ,
}
the propagators corresponding to different lattice sites form a complete orthonormal
basis on the \emph{whole} chain. The difference matrix in \eqref{DCK} is built by choosing
a subset $\ell \in X$ of this basis, and then reducing the matrix to the subsystem $m,n \in A$.
Now the duality amounts to interchanging the roles of the basis and spatial indices,
by defining the so-called overlap matrix
\eq{
M_{\ell \ell'}(t) = \sum_{m \in A} I_{m-\ell}^*(t)I_{m-\ell'}(t) \, ,
\label{A}
}
where the indices are restricted to $\ell, \ell' \in X$. It is then easy to show, that the
nonzero eigenvalues of the reduced matrices $\Delta C_A(t)$ and $M_X(t)$ are identical.
In fact, this is completely analogous to the duality between the reduced correlation
matrix of a Fermi sea ground state and the overlap matrix indexed by the occupied
momenta \cite{Klich06,CMV11,EP13}.

The duality relation is particularly useful if the number of holes is small compared to
the size of the subsystem. Indeed, the result \eqref{dzetat1h} for the single hole,
as well as $\eqref{dzetat2hb}$ for the double hole are immediately reproduced.
Furthermore, one could consider directly the limit of a half-infinite
chain, in which case the expression further simplifies to
\eq{
M_{\ell \ell'}(t) = i^{\ell-\ell'}\sum_{m=1}^{\infty}
J_{m-\ell}(t) J_{m-\ell'}(t)= i^{\ell-\ell'} B_{1-\ell,1-\ell'}(t) \, ,
\label{AB}}
where
\eq{
B_{mn}(t) = \frac{t}{2(m-n)} \left[J_{m-1}(t)J_{n}(t) - J_{m}(t) J_{n-1}(t) \right]
\label{B}}
is the so-called discrete Bessel kernel. This kernel is known to describe the time-evolved
correlation matrix for a step-like initial state (i.e. the limit $K\to\infty$),
also known as the domain-wall quench \cite{ARRS99,HRS04,AKR08,ER13,VSDH15,MNS19}.

We first consider the case where the excitation is given by a contiguous set of $K$ holes
located just next to the interval, $X=\left[-K+1,0\right]$. Then, by relation \eqref{AB}, we find that
the half-chain entanglement for an extended excitation is exactly equal to that
of a \emph{finite} interval $\left[1,K\right]$ in the domain-wall quench. In other words,
the size of the subsystem and that of the excitation are dual to each other. In particular,
the case $K \to \infty$ corresponding to the domain wall is self-dual.

%%%%%%%%%%%%%%%%%%%%%%%%%%%%%%%%%%%%%%%%%%%%%%%%%%%%%%%%%%%
%
\begin{figure}[htb]
\center
\includegraphics[width=0.6\textwidth]{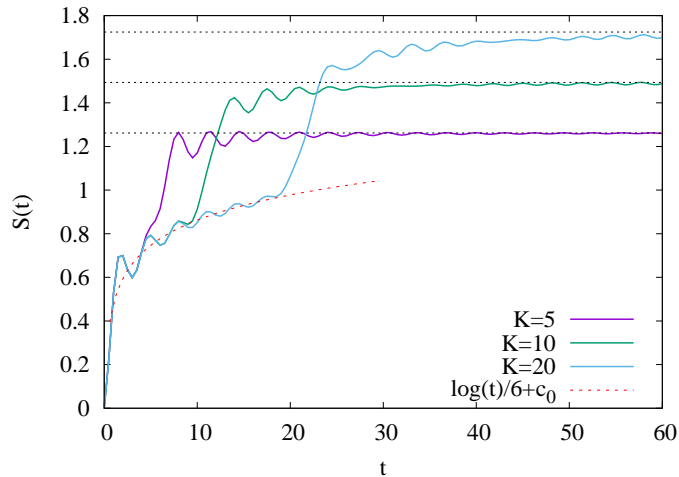}
\caption{Evolution of the half-chain entropy  for a contiguous set of $K$ holes at complete filling.
The red dashed line shows the domain-wall result \eqref{Sdw}. Horizontal dotted
lines indicate the ground-state entropy of an interval of length $K$ at half filling.}
\label{fig:entK}
\end{figure}
%
%%%%%%%%%%%%%%%%%%%%%%%%%%%%%%%%%%%%%%%%%%%%%%%%%%%%%%%%%%%

In Fig.~\ref{fig:entK} we show the results for the entropy for various $K$.
One can observe that up to $t \approx K$, the entropy follows the domain-wall
quench result which, up to oscillations, shows a logarithmic growth
\eq{
S_{dw} = \frac{1}{6} \ln (t) + c_0 \,
\label{Sdw}}
with $c_0 \approx 0.4785$ \cite{EIP09,EP14,DSVC17}. This is indicated by the red dashed line.
Around $t \approx K$, where the last part of the extended hole propagates
into the right half-chain, the entropy shows an abrupt jump. 
From here on, one can observe the convergence to a finite and $K$-dependent
value. Using the asymptotics of the discrete Bessel kernel
with $m,n$ fixed one has \cite{BOO00}
\eq{
\lim_{t \to \infty} B_{mn}(t) = \frac{\sin \left[\frac{\pi}{2}(m-n)\right]}{\pi(m-n)} \, ,
\label{sink}}
which is nothing else but the ground-state correlation matrix at half filling. Hence, we can
conclude that the asymptotic entropy assumes the ground-state value for an
interval of length $K$, shown by the horizontal dotted lines in Fig.~\ref{fig:entK}.

The interpretation of the result is the following. If the $K$ holes are located
at neighbouring sites, they effectively behave as a composite excitation.
Similarly to the case of a single hole, half of the density propagates to the
right and the other half to the left.
However, the coherence between parts of the composite excitation is preserved and,
despite the density being spread out over large distances, the inhomogeneous
half-infinite chain effectively behaves as a finite homogeneous one of size $K$.
The asymptotic entanglement thus scales as $1/3 \ln K$, i.e. the contribution
of the holes is not additive.

%%%%%%%%%%%%%%%%%%%%%%%%%%%%%%%%%%%%%%%%%%%%%%%%%%%%%%%%%%%
%
\begin{figure}[htb]
\center
\includegraphics[width=0.49\textwidth]{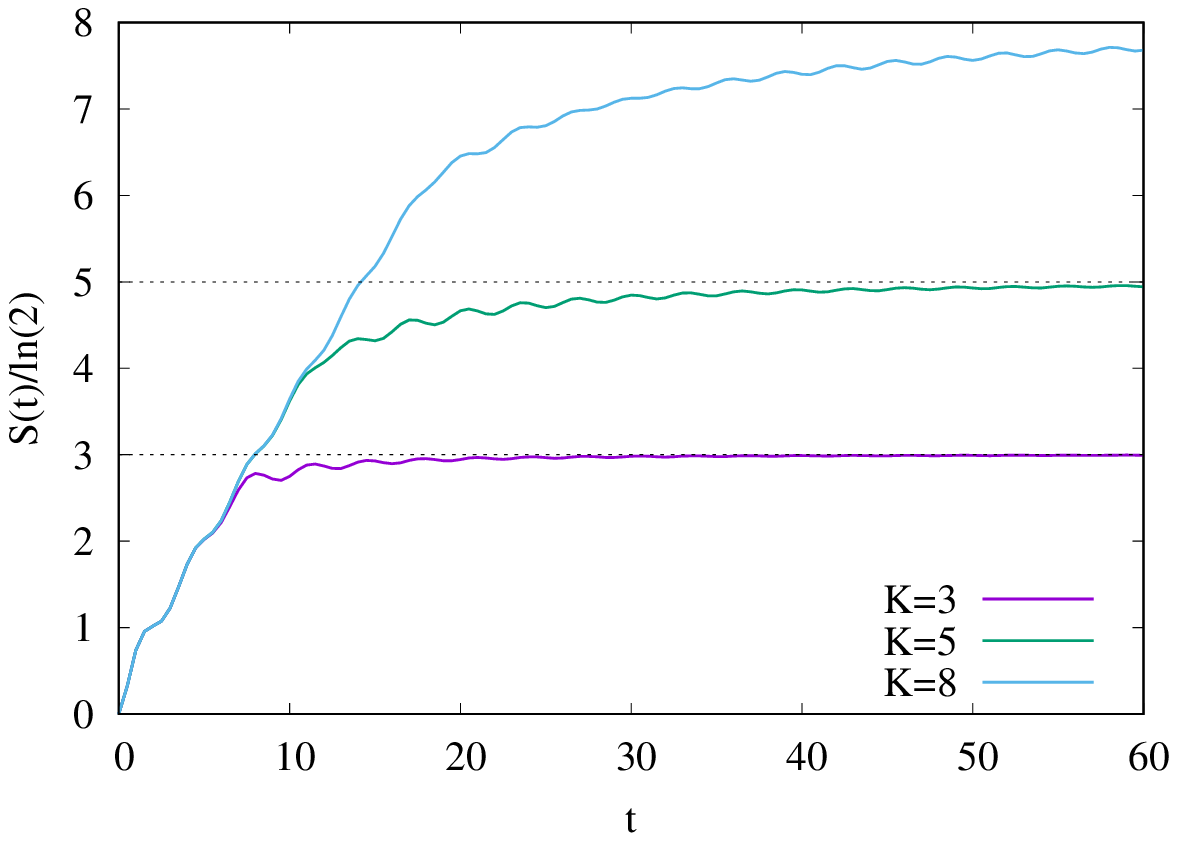}
\includegraphics[width=0.49\textwidth]{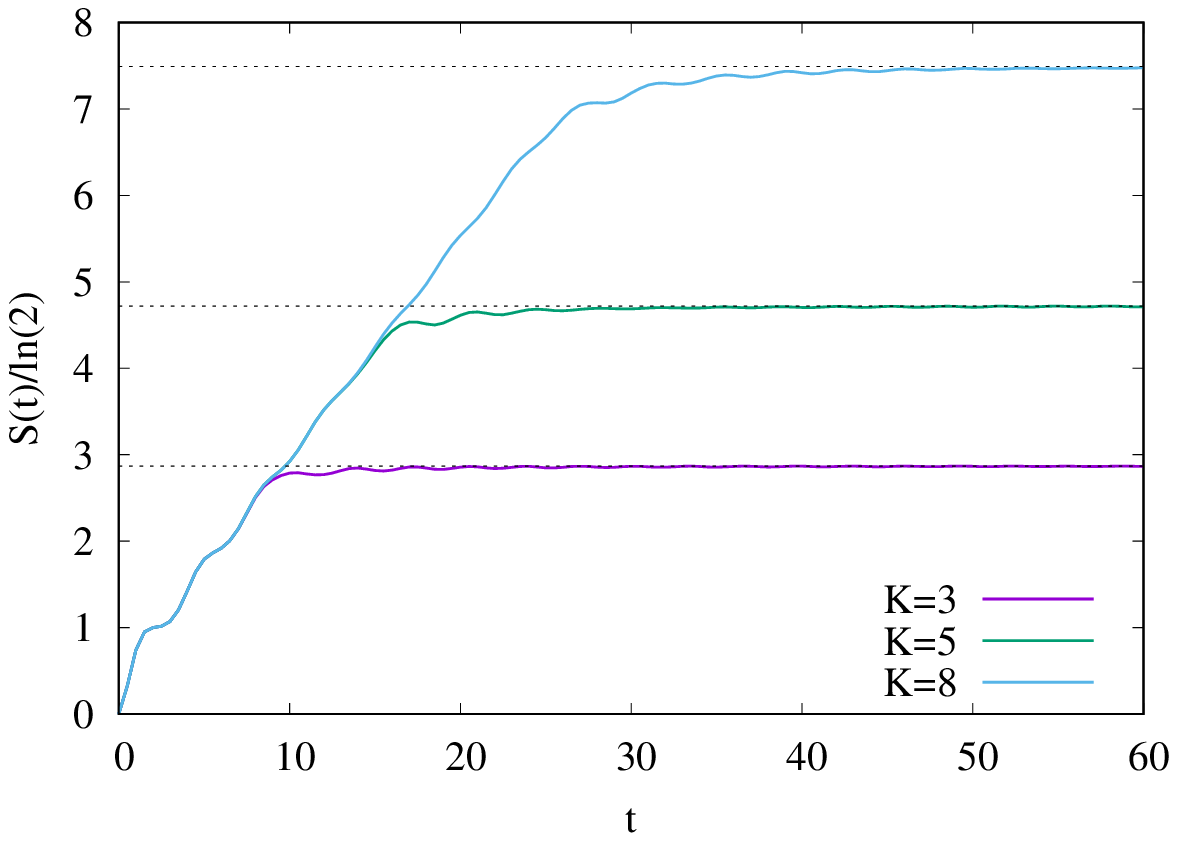}
\caption{Evolution of the half-chain entropy for a set of $K$ holes with
periodicity $p=2$ (left) and $p=3$ (right) in a completely filled chain.
Horizontal dotted lines indicate the ground-state entropy of a non-contiguous
subsystem with $K$ sites at equal distances $p$ at half filling.}
\label{fig:entKp}
\end{figure}
%
%%%%%%%%%%%%%%%%%%%%%%%%%%%%%%%%%%%%%%%%%%%%%%%%%%%%%%%%%%%

In the next example, we demonstrate what happens if the holes are located on
non-adjacent sites, i.e. they become distinguishable. For simplicity, we shall
check only the case where $K$ holes are placed periodically at sites $0, -p, -2p, \dots, -(K-1)p$.
In Fig.~\ref{fig:entKp} we show the cases $p=2$ (left) and $p=3$ (right), respectively.
One can immediately see that the situation is now completely different,
with the increase being linear and the asymptotic entropy proportional to $K$.
Indeed, due to the distinguishability of the excitations, they start to contribute
independently to the entropy, and one can recognize a kind of step structure
in the entropy increase. Regarding the asymptotic value, this is still given by the
discrete sine kernel \eqref{sink}, however with a subsystem that includes only
every $p$-th site. Such a non-contiguous subsystem is well known to have
an extensive entropy \cite{KMN06,IP10}. In particular, the case $p=2$ corresponds
to the situation where the asymptotic overlap matrix becomes diagonal, due to
the special checkerboard structure of the matrix elements in \eqref{sink}, and thus one has
$S(\infty)=K \ln 2$. In the case $p=3$, the subsystem involves odd distances,
thus preserving some of the correlations between the holes which leads
to $S(\infty) < K \ln 2$. The horizontal dotted lines indicate the actual values
of the asymptotic entropy, and one can see a clear convergence towards them.

\begin{figure}[htb]
\center
\includegraphics[width=0.49\textwidth]{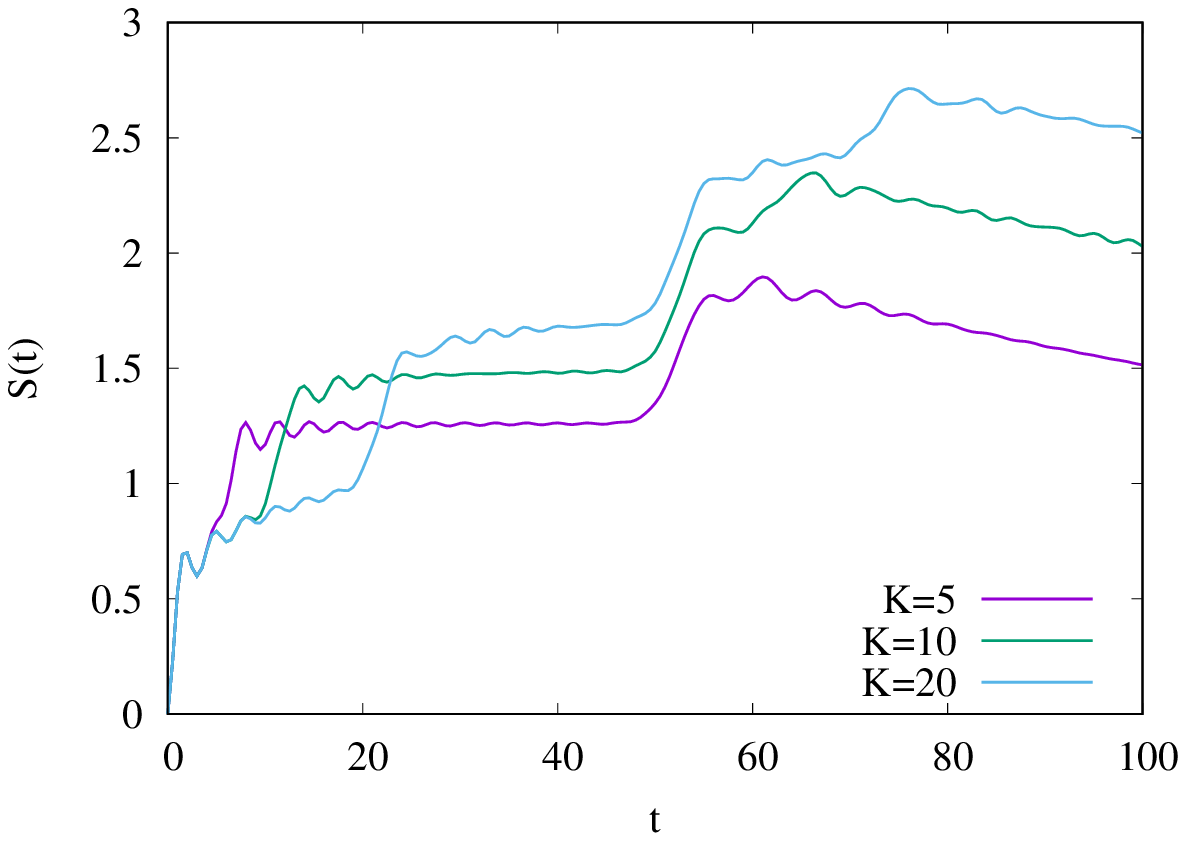}
\includegraphics[width=0.49\textwidth]{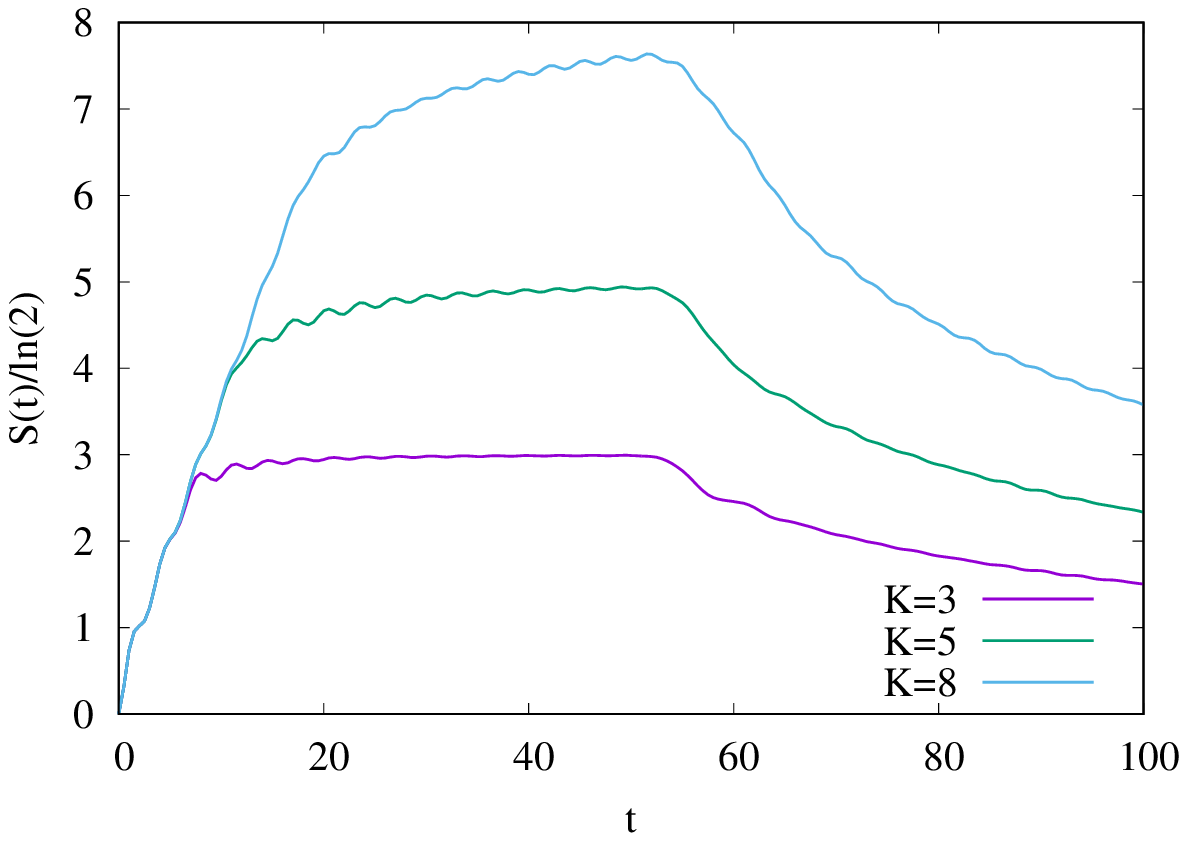}
\caption{Entropy evolution for a finite block of length $L=50$,
for a contiguous set of $K$ holes (left), as well as with a spacing $p=2$ (right)
in a completely filled chain. Note the different vertical scales.}
\label{fig:entKpfin}
\end{figure}

Finally one can check what happens when considering a finite subsystem,
in which case we have to use the form \eqref{A} of the overlap matrix.
The results for $L=50$ are shown in Fig.~\ref{fig:entKpfin}, for both a
contiguous set of holes (left) as well as for $p=2$ (right).
In the latter case the entropy starts to decrease immediately after
$t\approx L$, when the first hole leaves the subsystem, and the
decay continues in a monotonous fashion as the other holes follow.
In contrast, for the extended excitation one first observes a jump
at $t\approx L$, followed by a second smaller jump at $t\approx L+K$,
when the extended excitation leaves the interval. The slow decay of the
entropy sets in only after this transient, the quantitative description of which
would require some further investigations.

%%%%%%%%%%%%%%%%%%%%%%%%%%%%%%%%%%%%%%%%%%%%%%%%%%%%%%%%%%%
%
%
%%%%%%%%%%%%%%%%%%%%%%%%%%%%%%%%%%%%%%%%%%%%%%%%%%%%%%%%%%%

\section{Conclusions\label{sec:concl}}

We have studied the spreading of entanglement for a family of excitations that
can be written as a product of a few local fermion operators in a hopping chain.
In case of a single particle or hole, the excess entanglement is essentially
determined by the density fraction of the excitation located within the subsystem.
This can be well approximated by a semi-classical formula involving the
single-particle velocities of the fermionic modes. The result is thus completely
analogous to the ones found for local fermionic excitations in the XY \cite{EM18,EM20}
as well as the XXZ chain \cite{GE21}. The situation is more complicated for a double
hole excitation, where some of the coherence between the holes is preserved by the
time evolution. The two holes thus do not behave independently and the additivity of
the excess entropy is recovered only in the limit of large separations.
In fact, this behaviour is completely analogous to the case of two-particle eigenstates,
where the additivity is recovered only for large differences between the momenta \cite{ZR20a,ZR20b}.

In sharp contrast, the coherence of particle-hole excitations decays rapidly in time
and thus the entropy becomes additive for arbitrary separations.
It should be stressed, however, that it is crucial to consider a particle-hole excitation
that is orthogonal to the ground state. If the vacuum expectation value in \eqref{phi0ph}
is not subtracted, then offdiagonal contributions appear in the difference matrix, similarly
to the case of the double hole in \eqref{DC2h}. In the particle-hole case, however, we observed
that the eigenvalues $\Delta\zeta_{1,2}$ of the difference matrix cannot be directly
connected to the anomalous eigenvalues in the $\zeta_k(t)$ spectrum.
This is probably due to the nonvanishing overlap between the ground and excited states.
In any case, we found that the asymptotic excess entropy is given by a nontrivial value
that is strictly less then $2 \ln2$, converging towards it only for large separations.
Understanding this mechanism requires some further investigations.

The case of multiple hole excitations was studied only at complete filling,
and could be related to the domain-wall problem by a duality argument.
A contiguous set of holes was found to behave as a composite excitation,
with the asymptotic entropy scaling logarithmically with the number of holes.
On the other hand, a finite spacing between the holes makes them distinguishable,
and leads to a linear scaling of the entropy. In fact, this latter scenario is more
reminiscent to the one observed in a global or inhomogeneous quench \cite{CC05,FC08,AC17,ABFPR21}.
It would be interesting to check how these results for multiple excitations generalize to
arbitrary fillings.

The characteristic feature of a local excitation is the appearance of
an anomalous eigenvalue in the reduced correlation matrix spectrum,
which is responsible for the excess entropy. One should note that this behaviour
is very reminiscent of the one found earlier in the local quench problem, for a
subsystem chosen symmetrically around the initial cut \cite{EP07}. Indeed,
this mechanism was found to lead to the slow asymptotic decay \eqref{DSas1h} of the entropy
in both cases. However, while for a local excitation the anomalous eigenvalue is related
to the change of density in the subsystem, in a local quench the density is unchanged
and thus the origin of the mechanism must be more subtle. 
On the other hand, the half-chain entropy in the local quench shows a logarithmic
increase \cite{CC07,SD11}, in contrast to a finite excess entropy for the local operator excitation.
The proper understanding of the similarities and differences between these two nonequilibrium
protocols is an interesting open question for future research.

%Corrections to entropy due to noncommutativity

\ack
The author acknowledges funding from the Austrian Science Fund (FWF) through
project No. P30616-N36.

\appendix

\section{Proof of Gaussianity\label{app:a}}

In this Appendix we shall prove, following the lines of Ref. \cite{CR17},
Gaussianity for a large class of excitations that includes the ones in the main text.
For this purpose, let us consider a general quadratic Hamiltonian
\eq{
H= \frac{i}{4} \sum _{m, n=1}^{2N} H_{mn} a_m a_n
}
written in terms of Majorana fermions
\eq{
a_{2m-1}=c_m + c^\dag_m \, , \qquad a_{2m}=i(c_m - c^\dag_m ) \, .
}
We show that the state $\mathcal{O}\ket{\psi_0}$ is Gaussian when
\eq{
\mathcal{O} = \sum_m w_m a_m \, .
%\bra{\psi_0} \mathcal{O}^\dag\mathcal{O} \ket{\psi_0}=1
\label{O}}
In other words, acting with an arbitrary linear combination
of Majorana operators on the ground state leads to a Gaussian excited state.

We shall first consider the case where the coefficients $w_m$ are real,
thus the operator is unitary, $\mathcal{O}^\dag\mathcal{O}=\mathbb{1}$.
In this case the excited state is clearly Gaussian. Indeed, working in the
Heisenberg picture, one obtains a linear combination of the original operators
\eq{
a'_m = \mathcal{O}^\dag a_m \mathcal{O} = \sum_{n} A_{mn} a_n \, , \qquad
A_{mn} = 2w_mw_n-\delta_{mn} \, .
}
An arbitrary $p$-point correlation can then be rewritten as
\eq{
\braket{\mathcal{O}^\dag a_{m_1}a_{m_2}\dots a_{m_p}\mathcal{O}} =
\braket{a'_{m_1}a'_{m_2}\dots a'_{m_p}} = \sum_{n_1,n_2,\dots,n_p}
A_{m_1n_1}A_{m_2n_2} \dots A_{m_pn_p} \braket{a_{n_1}a_{n_2}\dots a_{n_p}}
}
and Gaussianity follows from the multilinearity of Wick's theorem.

The general case will be handled as follows. For an arbitrary operator
$\mathcal{O}$, we shall construct a unitary $U$ such that
$\mathcal{O}\ket{\psi_0}=U\ket{\psi_0}$ is satisfied, and thus the Heisenberg
operators can be defined via $a'_m = U^\dag a_m U$.
First note, that for an arbitrary quadratic Hamiltonian there exists
an orthogonal transformation
\eq{
a_m = \sum_k (R_{m,2k-1} d_{2k-1} + R_{m,2k} d_{2k})
}
that brings the skew-symmetric Hamiltonian matrix in the canonical block-diagonal form
\eq{
R^THR=\bigoplus_{k=1}^{N} \left(
\begin{array}{cc}
0 & -\epsilon_k \\
\epsilon_k & 0
\end{array}
\right).
}
Rewriting the Majorana fermions as
\eq{
d_{2k-1}=b_k + b^\dag_k \, , \qquad d_{2k}=i(b_k - b^\dag_k ) \, ,
}
one arrives at the diagonal form
\eq{
H=\sum_k \epsilon_k(b_k^\dag b_k-1/2) \, .
}
Importantly, since the $\epsilon_k$ are chosen to be all positive,
the ground state of $H$ is the vacuum of the $b_k$ operators.

Let us now consider an operator as in \eqref{O} with general complex
weights $w_m$. Its action on the ground state can be written as
\eq{
\mathcal{O}\ket{\psi_0}=
\sum_{m,k} w_m (R_{m,2k-1} d_{2k-1} + R_{m,2k} d_{2k}) \ket{\psi_0}=
%\sum_{m,k} w_m (R_{m,2k-1} -iR_{m,2k})b_k^\dag \ket{\psi_0} =
\sum_k (v_{2k-1}-iv_{2k})b_k^\dag \ket{\psi_0},
}
where we used $b_k  \ket{\psi_0} =0$ and defined the real weights
\eq{
v_{2k-1} = \mathrm{Re}\big[{\sum_m w_m (R_{m,2k-1} -iR_{m,2k})}\big] , \qquad
v_{2k} = -\mathrm{Im}\big[{\sum_m w_m (R_{m,2k-1} -iR_{m,2k})}\big] .
}
Now we can replace again the operators $b_k^\dag$
by either $d_{2k-1}$ or $id_{2k}$, respectively, which allows us to rewrite
\eq{
\mathcal{O}\ket{\psi_0}=
%U\ket{\psi_0}=
%\sum_k (v_{2k-1}-iv_{2k})b_k^\dag \ket{\psi_0}=
\sum_k (v_{2k-1} d_{2k-1} + v_{2k}d_{2k}) \ket{\psi_0} ,
}
which is now an excitation with purely real coefficients.
Hence, transforming back to the original Majorana variables, we obtain
a unitary operator defined as
\eq{
U = \sum_m \tilde w_m a_m \, , \qquad
\tilde w_m = \sum_k (v_{2k-1} R_{m,2k-1}+v_{2k} R_{m,2k}) \, .
%{\sum_k(v^2_{2k-1}+v^2_{2k})}
}

Let us now consider an operator of product form
$\mathcal{O}=\mathcal{O}_2\mathcal{O}_1$, where the procedure described above
can be iterated. Indeed, one can rewrite
\eq{
a''_m = \mathcal{O}_1^\dag \mathcal{O}_2^\dag a_m \mathcal{O}_2 \mathcal{O}_1
= (\mathcal{O}'_2)^\dag a'_m \mathcal{O}'_2
\label{a2m}}
where the Heisenberg operators are defined as
\eq{
a'_m = U_1^\dag a_m U_1 \, , \qquad
\mathcal{O}'_2 = U_1^\dag \mathcal{O}_2 U_1 \, ,
%=\sum_m w'_m a'_m
}
and $U_1$ is the unitary associated to $\mathcal{O}_1$.
In the next step the Heisenberg operator $\mathcal{O}'_2$ is
written as a linear combination of the $a'_m$ operators, and
one follows the exact same procedure to find the unitary operator
$U'_2$ associated to it. Clearly, the procedure can be iterated
for an arbitrary number of such operators in a product form.

Finally, we show that the state
($\mathcal{O}_2\mathcal{O}_1-\alpha \mathbb{1})\ket{\psi_0}$
also remains Gaussian. As shown before, this state can be replaced
by $(U_2U_1-\alpha \mathbb{1})\ket{\psi_0}$, which is not any more a unitary excitation.
However, one has
\eq{
(U^\dag_1U^\dag_2-\alpha \mathbb{1}) (U_2U_1-\alpha \mathbb{1})=
(1+\alpha^2 -2 \alpha \sum_{m} \tilde w_{1,m} \tilde w_{2,m}) \, \mathbb{1}
}
where we have used
\eq{
U^\dag_1U^\dag_2+U_2U_1 =
\sum_{m,n} \tilde w_{1,m} \tilde w_{2,n} \left\{a_m,a_n\right\} ,
}
and the anticommutator gives $\left\{a_m,a_n\right\}=2\delta_{mn}$.
Hence, the unitarity of the operator $U_2U_1-\alpha \mathbb{1}$ can
be restored by a simple normalization.

\section{Stationary phase calculation\label{app:b}}

Here we derive the stationary phase approximation of \eqref{dzetat1h}.
We need to evaluate the following oscillatory double integral
\eq{
\Delta \zeta(t)=\frac{\pi}{q_F} \sum_{n=1}^{L}
\int_{-q_F}^{q_F} \frac{\dd p}{2\pi} \int_{-q_F}^{q_F} \frac{\dd q}{2\pi}
\ee^{-it\cos p} \ee^{-ip(n-\ell)} \ee^{it\cos q} \ee^{iq(n-\ell)} .
}
It is easy to see, that the stationary phase condition yields $q=p$,
one can thus introduce the variables $Q=q-p$ and $P=(q+p)/2$ and
expand the phases around $Q=0$. Using
\eq{
\sum_{n=1}^{L} \ee^{iQ(n-\ell)}= 
\ee^{iQ(1-\ell)}\frac{1-\ee^{iQL}}{1-\ee^{iQ}} ,
}
and extending the integral over $Q$ up to infinity, one has
\eq{
\Delta \zeta(t) \simeq
\int_{-q_F}^{q_F} \frac{\dd P}{2q_F} \int_{-\infty}^{\infty} \frac{\dd Q}{2\pi i}
\frac{\ee^{iQ(L+1/2-\ell)}-\ee^{iQ(1/2-\ell)}}{2 \sin(Q/2)} 
\ee^{-it \sin(P)Q}\, .
}
Finally, using the integral representation of the Heaviside theta function
\eq{
\Theta(x) = \lim_{\epsilon \to 0} \int_{-\infty}^{\infty} \frac{\dd Q}{2\pi i}
\frac{\ee^{iQx}}{Q-i\epsilon} \, ,
\label{ht}
}
and setting $v_P=\sin(P)$, one arrives at
\eq{
\Delta \zeta(t) \simeq
\int_{-q_F}^{q_F} \frac{\dd P}{2q_F}
\left[\Theta(L+1/2-\ell-v_Pt)-\Theta(1/2-\ell-v_Pt)\right] .
}
Using the identity $\Theta(b-x)-\Theta(a-x)=\Theta(x-a)\Theta(b-x)$ for $a<b$
and considering only $\ell \le 0$, one obtains \eqref{dzetatsc} in the main text.

In a similar fashion one can also consider the approximation of the overlap
\eq{
\Delta = \frac{\pi}{q_F} \sum_{n=1}^{L}
\int_{-q_F}^{q_F} \frac{\dd p}{2\pi} \int_{-q_F}^{q_F} \frac{\dd q}{2\pi}
\ee^{-it\cos p} \ee^{-ip(n-\ell_2)} \ee^{it\cos q} \ee^{iq(n-\ell_1)} .
}
The only difference is the phase factor
\eq{
\ee^{-iq\ell_1+ip\ell_2} = \ee^{iP(\ell_2-\ell_1)} \ee^{-iQ(\ell_1+\ell_2)/2} \, ,
}
which then yields the semi-classical result
\eq{
\Delta \simeq
\int_{-q_F}^{q_F} \frac{\dd P}{2q_F} \ee^{iP(\ell_2-\ell_1)}
\left[\Theta(L+d-v_Pt)-\Theta(d-v_Pt)\right] , \qquad
d = \frac{1-\ell_1-\ell_2}{2} \, .
}

\section{Diagonalization of $\Delta C(t)$ for the double-hole excitation\label{app:c}}

According to \eqref{DC2h}, the problem can be cast in the following way
\eq{
\Delta C (t) = 
a \left(\ket{\psi_1}\bra{\psi_1} + \ket{\psi_2}\bra{\psi_2}\right)+
b \left(\ket{\psi_1}\bra{\psi_2} + \ket{\psi_2}\bra{\psi_1}\right) \, .
\label{DCapp}}
Clearly, however, the basis appearing in \eqref{DCapp} is not orthonormal.
Indeed, one has
\eq{
\lambda_\alpha  = \braket{\psi_\alpha | \psi_\alpha}= \sum_{n=1}^{L}|I_{n-\ell_\alpha}(t)|^2, \qquad
\Delta = \braket{\psi_1 | \psi_2}= \sum_{n=1}^{L} I^*_{n-\ell_2}(t) I^{\phantom{*}}_{n-\ell_1}(t) \, .
\label{lamdelapp}}
We thus define a new orthonormal basis
\eq{
\ket{\psi_1} = \alpha_1 \ket{\phi_1} + \alpha_2 \ket{\phi_2} , \qquad
\ket{\psi_2} = \beta_1 \ket{\phi_1} + \beta_2 \ket{\phi_2} ,
}
such that $\braket{\phi_\alpha|\phi_\beta}=\delta_{\alpha,\beta}$.
Substituting into \eqref{lamdelapp}, one obtains the following equations
\eq{
\lambda_1 = |\alpha_1|^2 + |\alpha_2|^2, \qquad
\lambda_2 = |\beta_1|^2 + |\beta_2|^2, \qquad
\Delta = \alpha_1^* \beta_1 + \alpha_2^* \beta_2 \, .
}
The remaining equation follows from the criterium that the matrix
$\Delta C(t)$ should be diagonal in the new basis. This gives
\eq{
a(\alpha_1 \alpha_2^* + \beta_1 \beta_2^*)+
b(\alpha_1 \beta_2^* + \beta_1 \alpha_2^*) = 0 \, ,
\label{ab0}}
whereas for the diagonal entries one has
\eq{
\Delta\zeta_1 = a(|\alpha_1|^2 + |\beta_1|^2) +
b ( \alpha_1^* \beta_1 + \beta_1^* \alpha_1) \, , \quad
\Delta\zeta_2 = a(|\alpha_2|^2 + |\beta_2|^2) +
b ( \alpha_2^* \beta_2 + \beta_2^* \alpha_2) \, .
}
The latter equations can be rewritten as
\eq{
\Delta\zeta_1 = a (\lambda_1 + x)  + b \, y \, , \qquad
\Delta\zeta_2 = a (\lambda_2 - x) + b(\Delta + \Delta^* - y)
\label{zeta12app}}
where the following new variables have been introduced
\eq{
x= |\beta_1|^2-|\alpha_2|^2 , \qquad
y = \alpha_1^* \beta_1 + \beta_1^* \alpha_1 \, .
}
By appropriate manipulations of \eqref{ab0} and its complex conjugate,
the following two equations can be obtained for the unknown variables
\eq{
a \left[|\Delta|^2 - x(x+\lambda_1-\lambda_2)\right] +
\frac{b}{2} \left[ y(\lambda_2-x)+ (\Delta+\Delta^*-y)(\lambda_1+x)\right]=0
}
\eq{
\begin{split}
a^2 \left[|\Delta|^2 - x(x+\lambda_1-\lambda_2)\right] &+
ab \left[ y(\lambda_2-x)+ (\Delta+\Delta^*-y)(\lambda_1+x)\right] \\
&+ b^2\left[ y(\Delta+\Delta^*-y)+\lambda_1\lambda_2 - |\Delta|^2\right]
=0 \label{xy}
\end{split}}
The sum of the eigenvalues follows immediately from \eqref{zeta12app} as
\eq{
\Delta\zeta_1 + \Delta\zeta_2 = a(\lambda_1+\lambda_2) + b(\Delta + \Delta^*) \, ,
}
whereas their product can be found, by using additionally \eqref{xy}, as
\eq{
\Delta\zeta_1 \Delta\zeta_2 = (a^2-b^2)(\lambda_1\lambda_2-|\Delta|^2) \, .
}
The above two equations can be combined to find the eigenvalues in \eqref{dzetat2h}.

\section*{References}

\bibliographystyle{iopart-num}

\bibliography{xxlocalex_refs}

\end{document}